\documentclass[11pt,a4paper]{article}

\usepackage{grffile}
\usepackage{graphicx}
\usepackage{amsmath,amsthm,amssymb,bm,amsfonts}
\usepackage{slashed}

\usepackage{url}
\usepackage{multirow}

\usepackage[utf8]{inputenc}

\RequirePackage[numbers,sort&compress]{natbib}

\usepackage{color}
\usepackage[normalem]{ulem}
\def\half{{\textstyle{\frac{1}{2}}}}

\def\qbar{{\overline{q}}}

\def\eto{{\rm{e}}}

\def\ftilde{\tilde{f}}

\def\ibar{{\bar{i}}}
\def\xbar{{\bar{x}}}

\def\fbar{{\bar{f}}}

\def\kperp{{k_T}}

\def\kperp{k_T}

\def\qperp{q_{T}}

\def\kperpone{k_{1T}}
\def\kperptwo{k_{2T}}

\newcommand{\be}{\begin{eqnarray}}
\newcommand{\ee}{\end{eqnarray}}

\newcommand{\bea}{\begin{align}}
\newcommand{\eea}{\end{align}}

\def\kvec{{\vec {k}}}
\def\qvec{{\vec{q}}} 
\def\pvec{{\vec{p}}}
\def\bvec{{\vec{b}}}
\def\eto{{e}}

\def\kperp{{k_T}}
\def\xbar{\bar{x}}

\def\fbar{\bar{f}}
\def\fbarbar{\bar{\fbar}}

\def\Pbar{\bar{P}}

\def\qvecperp{\vec{q}_T}
\def\kvecperp{\vec{k}_T}
\def\kvecperpone{\vec{k}_{1T}}
\def\kvecperptwo{\vec{k}_{2T}}

\def\kslash{\slashed{k}}

\def\Pslash{\slashed{P}}

\begin{document}

\begin{flushright}
\end{flushright}
\vspace{1cm}

\begin{center}
{\Large\textbf{Drell-Yan production with the CCFM-K evolution \\ \mbox{}}}\\
\vspace{.7cm}
Krzysztof Golec-Biernat$^{a,}$\footnote{\texttt{golec@ifj.edu.pl}}, Tomasz Stebel$^{b,}$\footnote{\texttt{tomasz.stebel@uj.edu.pl}}

\vspace{.3cm}
\textit{
$^a$Institute of Nuclear Physics PAN, Radzikowskiego 152, 31-342 Krak\'ow, Poland\\
$^b$ Institute of Physics, Jagiellonian University, S.\L{}ojasiewicza 11, 30-348 Krak\'ow, Poland
}
\end{center} 

\vspace*{2cm}
\begin{abstract}
We discuss the Drell-Yan dilepton production using the transverse momentum dependent parton distributions evolved
with the Catani-Ciafaloni-Fiorani-Marche\-sini-Kwieci\'nski (CCFM-K) equations in the single loop approximation. Such equations
are obtained assuming angular ordering of emitted partons (coherence) for $x\sim 1$ and transverse momentum ordering for $x \ll 1$.
This evolution scheme also contains the Collins-Soper-Sterman (CSS) soft gluon resummation. 
We make a comparison with a broad class of data on transverse momentum spectra of low mass Drell-Yan 
dileptons.

\end{abstract}

\clearpage
\tableofcontents
\setcounter{footnote}{0}

\section{Introduction}

The Drell-Yan (DY) dilepton production \cite{Drell:1970wh} is one of the most intensively studied processes in particle physics. The existence of a hard electroweak probe (photon or $Z$ boson), which doesn't interact strongly and decays into a pair of leptons, provides a clear experimental signature of the partonic interactions in the colliding hadrons and significantly simplifies their theoretical description. For these reasons, the DY process is very efficient tool for the investigation of hadronic structure \cite{Peng:2014hta}, in particular the distributions of partons' transverse momentum. The key concept of the theoretical analysis of DY scattering, called factorization, is a separation between long-range and short-range degrees of freedom. The basic, collinear factorization theorem assumes no transverse momenta of partons in hadrons \cite{Collins:1989gx}. Its application to the Drell-Yan production is very well established and commonly used. The present state of art calculation for the DY process includes next-to-next-to-leading order (NNLO) QCD corrections \cite{Ridder:2016nkl,Gehrmann-DeRidder:2016jns,Gauld:2017tww}.

There are several kinematical regimes where the fixed-order collinear QCD doesn't provide good description of data. In particular, when transverse momentum of lepton pair is much smaller than it's invariant mass, $q_T\ll M$, the large logarithms $\log^n (M/q_T)$ occur in all orders of perturbative expansion. These corrections are effectively resumed within the Collins, Soper and Sterman (CSS) approach \cite{Collins:1984kg}. As a result, the collinear factorization need to be replaced by a new factorization based on the transverse momentum distributions (TMDs) \cite{Angeles-Martinez:2015sea}. For state of art TMD analyses of Drell-Yan process, see \cite{Bacchetta:2017gcc,Bertone:2019nxa,Bizon:2019zgf,Camarda:2019zyx, Bacchetta:2019sam,Scimemi:2019cmh}.
It is interesting to ask what is the transition from the small transverse momentum region to the region of $q_T\sim M$, where fixed order perturbative QCD should apply. As it was shown in \cite{Bacchetta:2019tcu}, when one considers moderate values of $M\sim 5-19\,{\rm GeV}$, the fixed-order predictions underestimate data in the region of $q_T\sim M$. 

Such apparent troubles with the collinear factorization prompts us to approach the DY process using 
more general concepts like $k_T$-factorization \cite{Gribov:1984tu,Kuraev:1977fs,Balitsky:1978ic,Lipatov:1996ts,Catani:1990eg,Catani:1994sq} which allow to address the issue of intrinsic transverse momentum of partons. It should be mentioned, however, that
the $k_T$-factorization approach is less popular since the higher QCD corrections are much harder to obtain than in the collinear framework. Also, unlike collinear factorization, it lacks formal proof. Despite these facts, many theoretical and phenomenological analyses of the DY process were made \cite{Brodsky:1996nj,Kopeliovich:2000fb,Gelis:2002fw, Baranov:2008hj, Deak:2008ky, GolecBiernat:2010de,Hautmann:2012sh,Nefedov:2012cq, Motyka:2014lya,Basso:2015pba,Schafer:2016qmk,Motyka:2016lta,Brzeminski:2016lwh,Golec-Biernat:2018kem,Nefedov:2018vyt,Blanco:2019qbm}. The theoretical improvement was accompanied by many experimental results from fixed--target \cite{Ito:1980ev,Moreno:1990sf,McGaughey:1994dx,Webb:2003bj} and collider \cite{Antreasyan:1981eg,Aad:2014qja, CMS:2014jea,Aaij:2015gna,Aidala:2018ajl} experiments.

In this paper, we examine in detail the low mass DY productions using the $k_T$-factorization approach combined 
with the transverse momentum parton distributions defined in the Catani-Ciafaloni Fiorani-Marchesini 
branching scheme (CCFM) \cite{Ciafaloni:1987ur, Catani:1989sg,Catani:1989yc,Marchesini:1994wr}. The original motivation for the CCFM branching was to extend angular ordering (coherence) of soft parton emission in the space-like branching from the region of $x\sim 1$ to the region of small $x\ll 1$. In this way, a unified evolution equation for transverse momentum dependent gluon distribution was found with angular ordering in both regions of $x$ in the approximation called {\it all loop}. In the fully inclusive case, this equation interpolates between the DGLAP equation at moderate $x$ and the BFKL equation obtained in the small $x$ limit. 
It is worth emphasizing that the CCFM branching scheme for $x\sim 1$ contains the CSS resummation of soft gluon emissions \cite{Catani:1990rr}. 
The Monte Carlo implementation of the all loop CCFM branching scheme was done in \cite{Webber:1990rn,Jung:2000hk,Hautmann:2014uua}. 
A general Monte Carlo scheme for QCD evolution was also constructed with the Parton Branching method \cite{Hautmann:2017xtx,Hautmann:2017fcj,Martinez:2018jxt} and subsequent analyses were presented in \cite{Martinez:2019mwt,Martinez:2020fzs}.

In the region of large or moderate values of $x$, when the small $x$ coherence can be neglected, the CCFM
scheme gives the DGLAP evolution equation with coherence at large $x$ only. This approximation, called 
{\it single loop}, was studied in \cite{Webber:1990rn,Marchesini:1994wr} for the gluon distribution. The extension of this scheme in terms of evolution equations for both quark and gluon distributions was proposed by Kwieci\'nski in \cite{Kwiecinski:2002bx}. This is why we call them the CCFM-K equations. The parton distributions which are obtained by solving these equations depend on transverse momenta and their properties were analyzed in \cite{Gawron:2002kc, Gawron:2003qg,Gawron:2003np,RuizArriola:2004ui,Broniowski:2017gfp}.
The first analysis of the weak gauge boson production with the CCFM-K equations was done in 
\cite{Kwiecinski:2003fu} while the low mass DY production with photon was addressed in \cite{Szczurek:2008ga}. 
Similarly to the all-loop case, the one-loop CCFM-K equations contain a part of the CSS resummation of soft parton emissions \cite{Gawron:2003np}.

The main goal of the presented analysis is a comprehensive analysis of all available data on transverse momentum spectra in low mass DY production with the CCFM-K evolved parton distributions and the leading order cross sections computed in $k_T$ factorization. We assume the most economical form of the initial conditions for the evolution with only one adjustable parameter. In this way, we concentrate on the most important effects of the CCFM-K evolution which are responsible for good description of data for $q_T\sim M$. The small $q_T$ description is acceptable in most cases, especially for higher masses, although 
precise comparison should involve more adjustable parameters in the initial conditions like in the CSS approach.
This is left for future studies.

The paper is organized as follows. In section \ref{sec:2} we give an overview of the CCFM framework and its version proposed by Kwieci\'nski in which quark and gluon transverse momentum dependent distributions and the CCFM-K evolution equations are introduced. We also discuss the relation of the CCFM-K approach to the CSS formalism (with the full derivation presented in Appendix A).
In section \ref{sect_DYxs} we describe the application of the discussed formalisms to the leading order DY cross section with both on-shell and off-shell matrix elements in the CCFM-K case.
In section \ref{sect_comp_data} we show numerical results and compare them with the low mass DY data. We summarize in section \ref{sect_summary}. 

\section{CCFM approach}
\label{sec:2}

\subsection{Branching kinematics}
\begin{figure}[t]
\begin{center}
\includegraphics[width=0.18\textwidth]{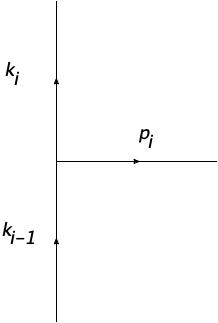}
\end{center}
\caption{Parton branching momenta}
\label{fig:1}
\end{figure}

Below we describe kinematics of the CCFM parton branching schemes in two approximations - single and all loop.
We work in the Sudakov base with two light cone vectors 
\be
P_1=\half\sqrt{S}(1,0,0,1)\,~~~~~~~~~~~~~~~~~~P_2=\half\sqrt{S}(1,0,0,-1)
\ee
in which the momenta in the branching process shown in Fig.~\ref{fig:1} are given by
\be
k_{i-1} =x_{i-1}P_1+\xbar_{i-1}P_2+k_{(i-1) T} 
\,,~~~~~~~~~~~~~~~~~
k_i = x_i P_1+\xbar_i P_2+k_{iT}.
\ee
Notice that the momentum fraction are proportional to the plus/minus components, e.g.
\be
x_i=\frac{k_i^+}{P_1^+} = \frac{k_{i}^0 +k_i^3}{\sqrt{S}}
\,,~~~~~~~~~~~~~~
\xbar_i=\frac{k_i^-}{P_2^-} = \frac{k_{i}^0 -k_i^3}{\sqrt{S}}. 
\ee
The emitted parton momentum can be found from the momentum conservation
\be
p_i = k_{(i-1) T}-k_{iT}= (x_{i-1}-x_i)P_1+(\xbar_{i-1}-\xbar_{i})P_2+(k_{(i-1) T}-k_{iT}),
\ee
where $x_{i-1}>x_i$. Denoting the transverse component by $p_{iT}=k_{(i-1) T}-k_{iT} =(0,\pvec_{iT},0)$
and assuming that $p_i^2=0$, one can compute the minus component to find 
\be
p_i=(x_{i-1}-x_i)\,P_1 +\frac{\pvec_{iT}^{\,2}}{(x_{i-1}-x_{i})S}\,P_2+p_{iT}
\ee
The rapidity of the emitted parton is given by
\be
y_i=\frac{1}{2}\ln\left(\frac{p_i^+}{p_i^-}\right) = -\ln\frac{|\pvec_{iT}|}{x_{i-1}(1-z_i)\sqrt{S}} ,
\ee
where $z_i={x_i}/{x_{i-1}}<1$. In the massless case $y_i=-\ln\tan({\theta_i}/{2})$ where $\theta_i$ is the emission angle with respect to the
$z$ axis defined by the collinear momenta $P_1$ and $P_2$, therefore
\be\label{eq:2.5a}
\tan\frac{\theta_i}{2} = \frac{|\pvec_{iT}|}{x_{i-1}(1-z_i)\sqrt{S}}\,.
\ee
The CCFM branching scheme \cite{Marchesini:1994wr} is defined with the help of the rescaled transverse momentum of the emitted parton,
\be\label{eq:2.7a}
\qvec_{iT}=\frac{\pvec_{iT}}{1-z_i}.
\ee
Thus, the transverse momentum conservation at the vertex $i$ reads
\be\label{eq:2.9}
\kvec_{(i-1)T} = \kvec_{iT} +(1-z_i)\, \qvec_{iT}.
\ee
From (\ref{eq:2.5a}) we obtain for the modulus
\be\label{eq:1.12}
q_{iT}\equiv |\qvec_{iT}|=x_{i-1}\sqrt{S}\tan({\theta_i}/{2}). 
\ee

The {\it single loop} approximation is defined by the condition
\be\label{eq:2.10a}
q_{iT} > q_{(i-1)T}\,,
\ee
thus for $z_i\to 0$ we find the transverse momentum ordering of {\it emitted} partons 
\be
|\pvec_{iT}|> |\pvec_{(i-1)T}|.
\ee
For finite $z\sim 1$, however, from (\ref{eq:1.12}) applied to (\ref{eq:2.10a}) we have
\be
x_{i-1}\tan\frac{\theta_i}{2} > x_{i-2}\tan\frac{\theta_{i-1}}{2}
\ee
and for $z_{i-1}=(x_{i-1}/x_{i-2})\to 1$
we obtain angular ordering of parton emissions
\be\label{eq:2.14a}
\theta_i > \theta_{i-1}.
\ee
Such a phenomenon is called coherence \cite{Ciafaloni:1987ur,Catani:1989sg,Catani:1989yc,Marchesini:1994wr}. 
Thus, in the single loop approximation partons are emitted with 
transverse momentum ordering for $z\to 0$ and angular ordering for $z\to 1$.
The {\it all loop} approximation is defined by the condition
\be\label{eq:2.2}
q_{iT} > z_{i-1}\,q_{(i-1)T}
\ee
which is equivalent to the condition
\be
\tan\frac{\theta_i}{2} > \tan\frac{\theta_{i-1}}{2}\,,
\ee
giving the angular ordering condition (\ref{eq:2.14a}). Thus, in the all loop approximation partons are emitted with 
angular ordering for any value of $z$.

\subsection{CCFM equation in all loop approximation}

The CCFM branching schemes allow to define the corresponding parton distributions. In the all loop approximation \cite{Marchesini:1994wr}
only the gluon distribution $f_g$ is defined up till now through the equation 
\begin{align}\nonumber
f_g(x,\kperp,Q) &= f_g^0(x,\kperp,Q_0)
+\int \frac{d^2\qvec}{\pi q^2}\int_x^1\frac{dz}{z}\,\theta(Q-zq)\theta(q-Q_0)\,\frac{\alpha_s(q)}{2\pi}
\Delta_S(Q,zq)\
\\ \label{eq:3.1}
&\times (2N_c)\left[\frac{\Delta_{NS}(\kperp,q,z)}{z}+\frac{\theta(1-z-Q_0/q)}{1-z}\right]
f_g\!\left(\frac{x}{z},|\kvec_T+(1-z)\qvec|, q\right)
\end{align}
which relates the gluon distribution at vertex $i$ with the gluon distribution at vertex $(i-1)$. In the above, 
$\kperp=|\kvec_T|$ and $q=|\qvec|$ are transverse momenta depicted in Fig.~\ref{fig:1} 
and $\Delta_S$ is the Sudakov form factor given by
\be
\Delta_S(Q,zq) =\exp\bigg\{-\int_{(zq)^2}^{Q^2}\frac{dp^2}{p^2}\frac{\alpha_s(p^2)}{2\pi}\int_0^{1-Q^0/p}dz^\prime z^\prime
P_{gg}(z^\prime)\bigg\},
\ee
where $P_{gg}$ is the gluon-gluon splitting function (\ref{eq:2.23a}), which resums virtual corrections for $z\to 1$.
For $z\to 0$, the virtual emissions are resumed by the non-Sudakov form factor
\be
\Delta_{NS}(\kperp,q,z) =\exp\bigg\{
-\int_{z}^{1}dz^\prime\,\frac{2N_c}{z^\prime}
\int^{k^2_T}_{(z^\prime q)^2}\frac{dp^2}{p^2} \frac{\alpha_s(p^2)}{2\pi}\bigg\}\,.
\ee
Notice that only the $1/z$ part of $P_{gg}(z)$ is present under the integral.
The first theta function in eq.~(\ref{eq:3.1}) reflects ordering (\ref{eq:2.2}) in which
$Q$ is a hard scale which terminates the CCFM evolution,
\be
zq=z\left(\frac{x}{z}\right)\sqrt{S}\tan\frac{\theta}{2}<Q\,.
\ee
Therefore, for given $x$ and $\sqrt{S}$, the hard scale determines the maximal emission angle
\be
\theta_{\rm max} = 2\arctan\frac{Q}{x\sqrt{S}}. 
\ee
The second theta function in (\ref{eq:3.1}) imposes the condition
\be
q>Q_0\gg \Lambda_{QCD}
\ee
which assures that $\alpha_s(q)\ll 1$ and the CCFM evolution scheme is perturbative. From the third theta function in (\ref{eq:3.1}), we find the following condition for the real gluon emission
\be
0<z<(1- {Q_0}/{q)}\,,
\ee
which allows to avoid singularity of $P_{gg}(z)$ at $z=1$.
Non-perturbative effects are encoded in the initial condition, $f^0_g(x,\kperp,Q_0)$, imposed at a scale $Q_0$. 

Eq.\,(\ref{eq:3.1}) can be used for the Monte Carlo generation of a parton cascade with angularly ordered emissions which leads to the gluon distribution $f_g$. Intensive studies with the CCFM-K equations in all-loop approximation were done using Monte Carlo 
program CASCADE \cite{Webber:1990rn,Jung:2000hk,Hautmann:2014uua}.

\subsection{CCFM-K evolution equations}

The mixing between the transverse and longitudinal variables in the theta function $\theta(Q-zq)$ prevents writing eq.\,(\ref{eq:3.1}) in the form of an evolution equation. However, this can be done in the single loop approximation in which the branching scheme leads to the CCFM-Kwieciński (CCFM-K) evolution equations for both quark and gluon distributions. The evolution scale is defined in such a case by the rescaled momentum $Q$.

In order to obtain the CCFM-K evolution equations in the single loop approximation, 
the branching conditions in eq.\,(\ref{eq:3.1}) are replaced by \cite{Kwiecinski:2002bx,Gawron:2002kc,Gawron:2003qg,Gawron:2003np} \begin{align}\nonumber
\Theta(Q-zq)~~ &\to~~ \Theta(Q-q),
\\
\Delta_{NS}(k_T,q,z)~~ &\to ~~ 1.
\end{align}
Thus, for $z\to 0$, the angular ordering is replaced by the transverse momentum ordering while for $z\to 1$ the angular ordering is still valid. In addition, quark splittings, $q\to qg$, $\bar{q}\to \bar{q}g$ and $g\to q\qbar$, are taken into account which allow to introduce quark distributions $f_{i=1,\ldots,2N_f}$ in addition to the gluon distribution $f_g$. In this way we obtain \cite{Gawron:2003np}
\begin{align}\nonumber
f_{i}(x,\kperp,Q) = f_i^0(x,\kperp) &+ \int_0^1 \frac{dz}{z}\int\frac{d^2\qvec}{\pi q^2}\frac{\alpha_s(q^2)}{2\pi}\theta(Q-q)\theta(q-Q_0)
\\ \nonumber
&\times
\bigg\{
\theta(z-x)\Big[P_{qq}(z) f_i\!\left(\frac{x}{z},k_T^\prime,q\right)+P_{qg}(z)f_g\!\left(\frac{x}{z},k_T^\prime,q\right)\Big]
\\\nonumber
&-z P_{qq}(z) f_i(x,\kperp,q)
\bigg\} ,
\\\nonumber
f_g(x,\kperp,Q) = f_g^0(x,\kperp) &+ \int_0^1 \frac{dz}{z}\int\frac{d^2\qvec}{\pi q^2}\frac{\alpha_s(q^2)}{2\pi}\theta(Q-q)\theta(q-Q_0)
\\\nonumber
&\times
\bigg\{
\theta(z-x)\Big[P_{gq}(z)\sum_{i=1}^{2N_f} f_i\!\left(\frac{x}{z},k_T^\prime,q\right)+P_{gg}(z)f_g\!\left(\frac{x}{z},k_T^\prime,q\right)\!\Big]
\\\label{eq:CCFM-K}
&- z\left[P_{gg}(z) +2N_fP_{qg}(z)\right]f_g(x,\kperp,q)
\bigg\} ,
\end{align}
where the argument of the parton distributions on the r.h.s. equals
\be
k_T^\prime=|\kvec_T+(1-z)\qvec|.
\ee
The one loop real emission splitting functions are given by\footnote{The celebrated "$+$" prescription is taken into account by the negative virtual emission terms in eqs.~(\ref{eq:CCFM-K}).}
\begin{align}\nonumber
P_{qq}(z) &= C_F\frac{1+z^2}{1-z}, 
\\\nonumber
P_{qg}(z) &=T_R\left\{z^2+(1-z)^2\right\},
\\\nonumber
P_{gq}(z) &=C_F\frac{1+(1-z)^2}{z},
\\\label{eq:2.23a}
P_{gg}(z) &=2C_A\left\{ \frac{z}{1-z}+\frac{1-z}{z} +z(1-z)\right\},
\end{align}
where $C_F=4/3$, $C_A=3$, $T_R=1/2$ and $N_f$ is the number of active flavours.
Notice that after integrating both sides of eqs.~(\ref{eq:CCFM-K}) over $\kvecperp$, the ordinary DGLAP equations are found for
the collinear quark and gluon distributions,
\be \label{eq:2.28aaa}
q_i(x,Q)=\int {d^2\kvec_T}\,f_{i}(x,\kperp,Q)\,,~~~~~~~~~~
g(x,Q)= \int {d^2\kvec_T}\,f_{g}(x,\kperp,Q)\,,
\ee
Eqs. (\ref{eq:CCFM-K}) can be written with the help of the Fourier transformation
\be\label{eq:2.28}
\ftilde_{i,g}(x,\bvec,Q) = \int {d^2\kvec_T}\,\eto^{i\kvec_T\cdot\bvec}f_{i,g}(x,\kperp,Q) 
\label{Fourier_transf}
\ee
which for $\bvec=0$ gives the PDFs (\ref{eq:2.28aaa}). 
Since the parton distributions depend on $\kperp=|\kvecperp|$, we perform the azimuthal angle integration with help of the relation
\be\label{eq:2.30a}
\eto^{i\kvecperp\cdot\bvec}=J_0(\kperp b)+2\sum_{n=1}^\infty i^nJ_n(\kperp b) \cos\phi
\ee
and obtain the parton distributions which depend on $b=|\bvec|$,
\be\label{eq:2.32zz}
\ftilde_{i,g}(x,b,Q) = \pi\int_0^\infty dk_T^2\,J_0(\kperp b)f_{i,g}(x,\kperp,Q)\,.
\ee 
Thus, taking the Fourier transform of both sides of eqs.\,(\ref{eq:CCFM-K}), we find the evolution equations which are diagonal in $b$:
\begin{align}\nonumber
\frac{\partial \ftilde_i(x,b,Q)}{\partial \ln Q^2} = &\frac{\alpha_s(Q^2)}{2\pi}\int_0^1\frac{dz}{z}
\Bigg\{
\theta(z-x) J_0((1-z)Qb) \Big[P_{qq}(z) \ftilde_i\!\left(\frac{x}{z},b,Q\right)
\\\nonumber
&+ P_{qg}(z)\ftilde_g\!\left(\frac{x}{z},b,Q\right)\Big] - zP_{qq}(z) \ftilde_i(x,b,Q) \Bigg\},
\\\nonumber
\frac{\partial \ftilde_g(x,b,Q)}{\partial \ln Q^2} = &\frac{\alpha_s(Q^2)}{2\pi}\int_0^1\frac{dz}{z}
\Bigg\{
\theta(z-x) J_0((1-z)Qb) \Big[P_{gq}(z) \sum_{i=1}^{2N_f}\ftilde_i\!\left(\frac{x}{z},b,Q\right)
\\\label{eq:4.8a}
&+ P_{gg}(z)\ftilde_g\!\left(\frac{x}{z},b,Q\right)\Big] - z\left[P_{gg}(z) +2N_fP_{qg}(z)\right]\ftilde_g(x,b,Q)
\Bigg\}.
\end{align}
These are the CCFM-K equations which we use in our forthcoming analysis. As expected, 
for $b=0$ we obtain the DGLAP evolution equations for the collinear PDFs (\ref{eq:2.28aaa}), i.e.
\be\label{eq:33}
\ftilde_{i}(x,b=0,Q)= q_i(x,Q)\,,~~~~~~~~~~~~ \ftilde_{g}(x,b=0,Q)= g(x,Q)
\ee

It should be emphasized that the studies with the CCFM-K equations were also done using the Parton Branching (PB) method for the construction of the TMD parton distributions \cite{Hautmann:2017xtx,Hautmann:2017fcj} which is based on Monte Carlo algorithms.
Recently, the low mass DY production was analyzed with this method in \cite{Martinez:2020fzs}. The main difference between our 
approach and the PB method, aside from technical issues, lies in the treatment of the strong coupling constant $\alpha_s$ in the CCFM-K equations. We keep it outside the integrals on the rhs of eq.~(\ref{eq:2.32zz}) with the scale given by the evolution variable
$Q$, whereas in the PB method $\alpha_s$ is inside the integrals over $z$ since it depends on the transverse momentum
of an emitted parton, $\kperp=(1-z) Q$. In such a case, a cutoff on the upper limit of $z$ is necessary to avoid the Landau pole in $\alpha_s(\kperp)$.

\subsection{Initial conditions and $b$-dependence}

\begin{figure}[t]
\begin{center}
\includegraphics[width=\textwidth]{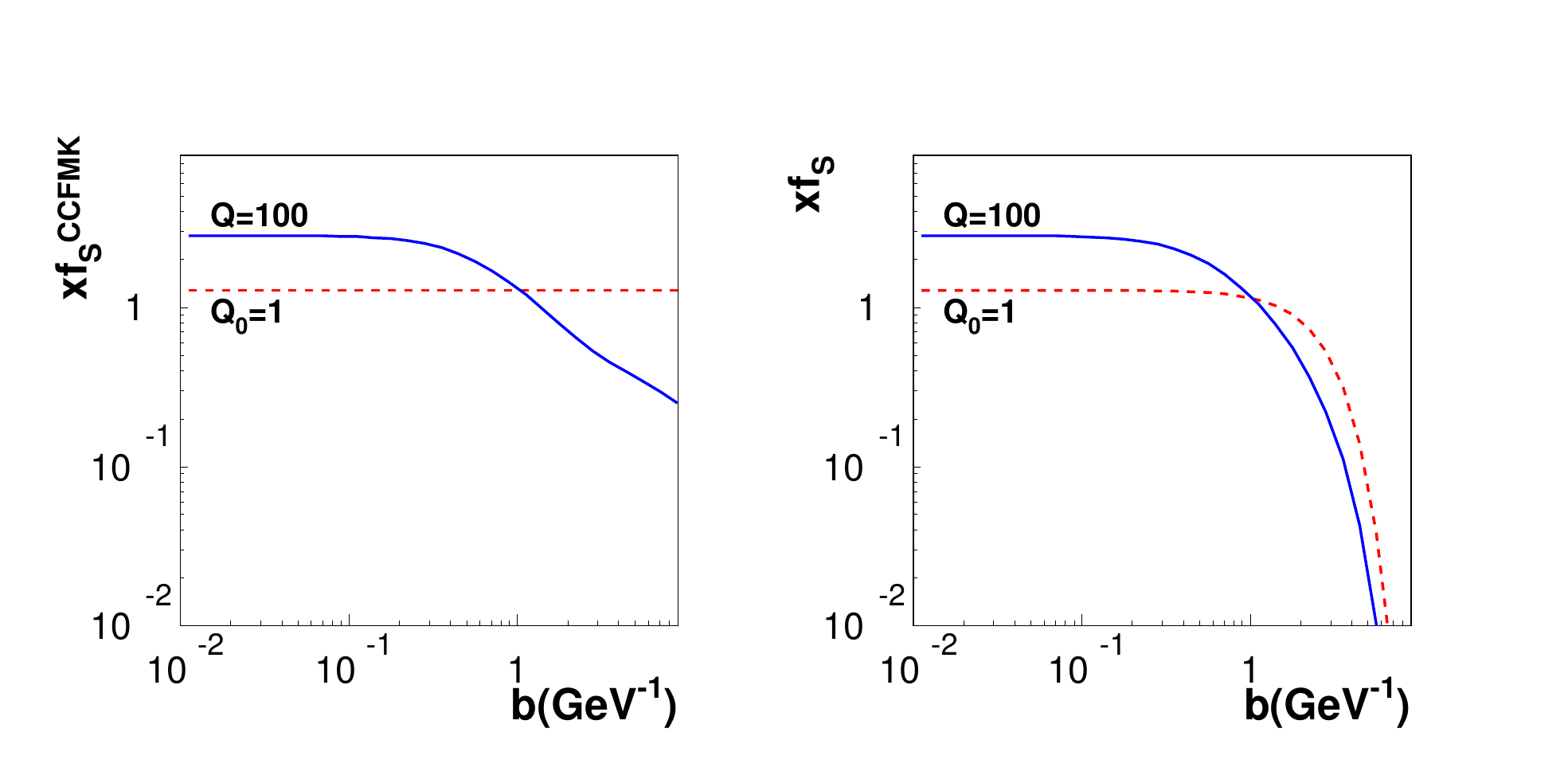}
\end{center}
\vskip -5mm
\caption{
The singlet quark distribution $f_S^{CCFMK}$ from eq.~(\ref{eq:2.35aa}) as a function of $b$ for fixed $x=10^{-2}$
and two scales: initial $Q_0=1~{\rm GeV}$ and final $Q=100~{\rm GeV}$ (left plot). These distributions where multiplied by the gaussian form factor (\ref{eq:2.37a}) on the right plot.
}
\label{fig:2}
\end{figure}

In order to solve eqs.~(\ref{eq:4.8a}), we need
initial conditions specified as functions of $x$ and $b$ at some perturbative scale $Q_0\gg \Lambda_{QCD}$.
They have to fulfill the conditions saying that for $b=0$ the collinear PDFs are recovered.
Thus the simplest possible choice is given in the factorized form 
\be\label{eq:4.10}
\ftilde_{i}(x,b,Q_0)= q_i(x,Q_0)F(b)\,,~~~~~~~~~~~~ \ftilde_{g}(x,b,Q_0)= g(x,Q_0)F(b),
\ee
where $q_i$ and $g$ are the LO collinear quark and gluon distributions at scale $Q_0$ 
and the non-perturbative form factor obeys the condition $F(0)=1$.
In the forthcoming analysis we will use the gaussian form factor with one free parameter $b_0$,
\be\label{eq:2.37a}
F(b) = \exp(-b^2/b_0^2).
\ee

In principle, different form factors can be used for quarks and gluons. However, with the common form factor, 
it is possible to write the solution of the CCFM-K equations for any value of $Q^2$ as a product
\be\label{eq:2.35aa}
f_{i,g}(x,b,Q)= f_{i,g}^{\rm CCFMK}(x,b,Q)\,F(b)\,.
\ee
where $f_{i,g}^{\rm CCFMK}$ is the solution for $F(b)\equiv 1$. This is because the equations (\ref{eq:4.8a})
are homogeneous, thus the multiplication by the common form factor $F(b)$ can be done at the beginning or the end of the evolution.
In this way, the perturbative and non-perturbative dependences of the solution are clearly separated.

This effect is shown in
Fig.\,\ref{fig:2} for the singlet quark distribution, $f_S=\sum_{i} f_i$, plotted as a function of $b$ for fixed $x=10^{-2}$.
The dashed curves are the initial conditions at $Q_0=1~{\rm GeV}$ with the MSTW08 LO PDFs \cite{Martin:2009iq}
while the solid curves are evolved to $Q=100~{\rm GeV}$. On the left plot, the $b$-dependence of the evolved curve is purely perturbative
while on the right plot the curves were multiplied by the form factor (\ref{eq:2.37a}). We see that its impact is the strongest for large values of $b$ while for small values, the $b$-dependence of the full solution remains perturbative.
After the Fourier transformation to the $\kperp$-space, we find broadening of the parton distributions due to the CCFM-K evolution, studied in detail in \cite{Gawron:2003qg,RuizArriola:2004ui,Broniowski:2017gfp}. 

\subsection{Relation to CSS resummation}

\begin{figure}[t]
\begin{center}
\includegraphics[width=\textwidth]{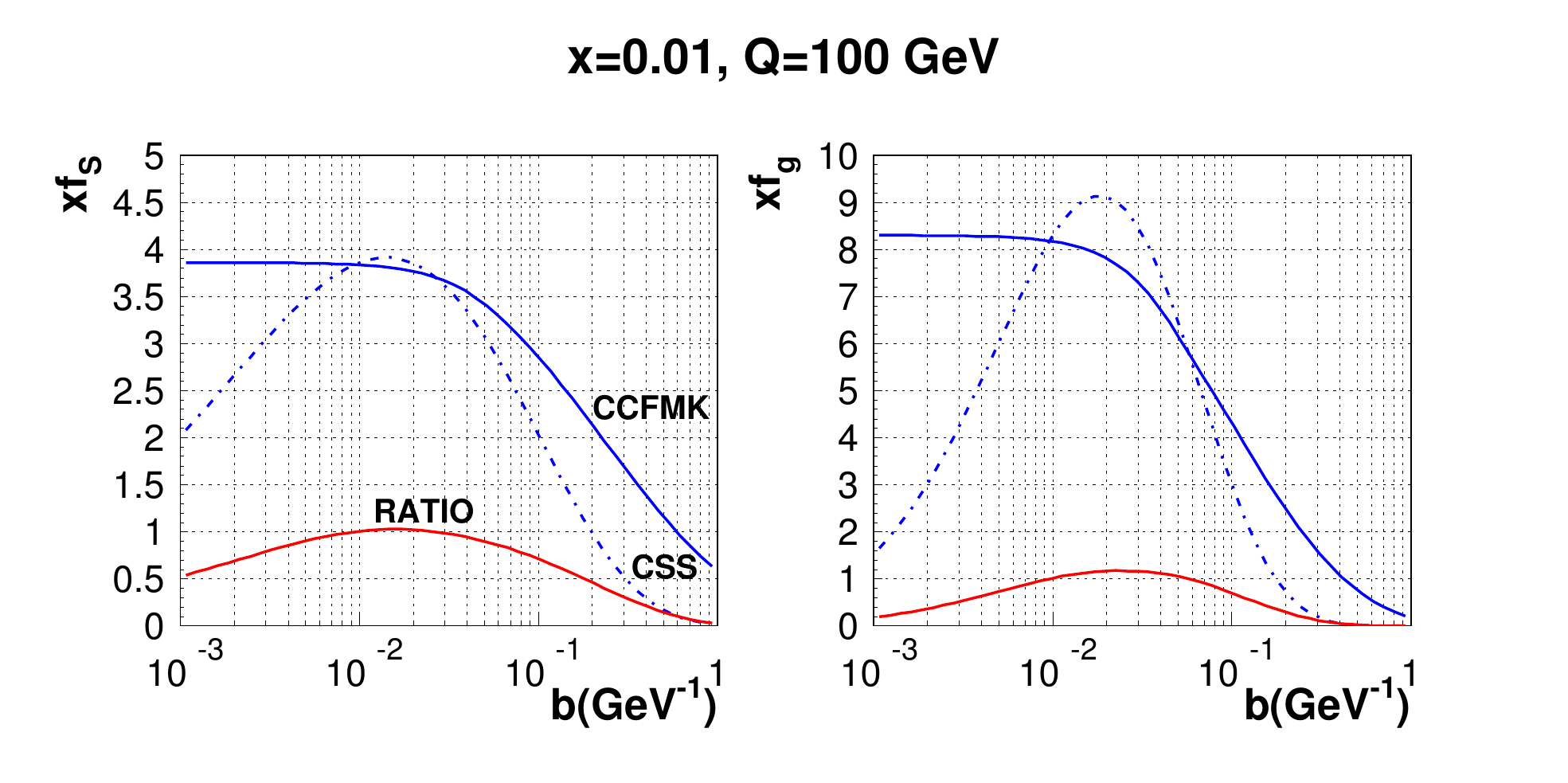}
\end{center}
\vskip -5mm
\caption{The solution of eqs.~(\ref{eq:4.8a}) for singlet quark (left plot) and gluon distributions (right plot) at $Q=100~{\rm GeV}$ and $x=10^{-2}$, obtained for $F(b)\equiv 1$ (solid lines) and the CSS approximation (\ref{eq:2.38ab}) (dot-dashed lines). The ratio 
CSS/CCFMK curves are shown as the red solid lines.
}
\label{fig:3}
\end{figure}

The CCFM-K equations contain a part of the Collins, Soper and Sterman (CSS) 
resummation of soft parton emissions in the limit $z\to1$. 
In Appendix \ref{app:A}, we present the proof that for the values of the parameter $b$ such that
\be\label{eq:2.39ab}
1/Q\ll b\ll 1/Q_0\,, 
\ee
the solution to the CCFM-K equations (\ref{eq:4.8a}) is given by the CSS formulas \cite{Collins:1984kg}
\begin{align}\nonumber
f_i(x,b,Q) &=\exp\bigg\{-\int_{1/b^2}^{Q^2}\frac{dq^2}{q^2}\frac{\alpha_s(q^2)}{2\pi}\left[A_{q}^{(1)}\ln(q^2b^2)+B_{q}^{(1)}\right]\bigg\}\,q_i(x,1/b)
\\\label{eq:2.38ab}
f_g(x,b,Q) &=\exp\bigg\{-\int_{1/b^2}^{Q^2}\frac{dq^2}{q^2}\frac{\alpha_s(q^2)}{2\pi}\left[A_{g}^{(1)}\ln(q^2b^2)+B_{g}^{(1)}\right]\bigg\}\,g(x,1/b)
\end{align} 
where the parameters $A_{q,g}^{(1)}$ and $B_{q,g}^{(1)}$ are defined in eq.\,(\ref{eq:a16}).
The above formulas were derived by picking large logarithms, $\ln(Qb)$ and $\ln(1/Q_0b)$, in the CCFM solution. It should be noted, however, that eqs.\,(\ref{eq:2.38ab}) do not contain the NLL (next-to-leading logarithmic) terms 
proportional to $\alpha_s^2$ under the integrals (see section \ref{DY:CSS}) since the splitting functions in eqs.~(\ref{eq:4.8a}) are in the leading order approximation.

In Fig.\,\ref{fig:3} we present the numerical solution to the CCFM-K equations (\ref{eq:4.8a}) with $F(b)=1$ (solid lines) against the CSS approximation (\ref{eq:2.38ab}) (dot-dashed lines) for the singlet quark (left plot) and gluon (right plot) distributions. For the chosen scales, $Q_0=1~{\rm GeV}$ and $Q=100~{\rm GeV}$, the range (\ref{eq:2.39ab}) corresponds to $b\in[10^{-2},1]$ GeV$^{-1}$.
We see that the CSS approximations extracted from the CCFM-K equation works reasonable well for
$b\in[10^{-2},10^{-1}]$~GeV$^{-1}$. For $b=10^{-2}$~GeV$^{-1}$ the two analyzed curves coincide, which 
results from the observation that for the scale $Q=1/b=100~{\rm GeV}$, corresponding to this point, both the CSS formulas (\ref{eq:2.38ab}) and the CCFM-K solution are equal to the collinear PDFs at the scale $Q$.
This is obvious for eqs.~(\ref{eq:2.38ab}), while for the CCFM-K solution it is a manifestation of the DGLAP limit (\ref{eq:33}) at $b=0$, which becomes already effective for $b=10^{-2}$~GeV$^{-1}$. Beyond the lower limit in (\ref{eq:2.39ab}),
the CSS approximation significantly deteriorates and the approximation (\ref{eq:2.38ab}) cannot describe the CCFM-K solutions.

The condition $1/Q< b$ which was necessary for us to find the connection between the CCFM-K and CSS approaches is not present in the original CSS formulation \cite{Collins:1984kg}, where $b$ can be arbitrary small. Nevertheless, recent studies \cite{Collins:2016hqq} introduces such a constraint, i.e. $b$ is limited from below by $b_{\rm min}\sim1/Q$. Analyzing this idea in the context of the CCFM-K approach, however, would go beyond the main thrust of our analysis.

\section{Drell-Yan cross section with $\kperp$-dependent PDFs}
\label{sect_DYxs}
The Drell-Yan cross section differential in photon's momentum is given by
\be\label{eq:3.1xxx}
\frac{d\sigma^{DY}}{dy_\gamma\,dM^2 \,d^2\qperp} =\frac{\alpha_{\rm em}^2}{24\pi^3S^2M^2}
\left(-W^{\mu}_{~\mu}\right),
\ee 
where $(y_\gamma, M,\qperp)$ are photon's rapidity, virtuality and transverse momentum while $W^{\mu}_{~\mu}$ is the trace of the hadronic tensor $W^{\mu\nu}$. 
With the lowest order matrix element for the process $q\qbar\to \gamma^*$, the trace is given by the transverse momentum factorization formula
\begin{align}\nonumber
W^{\mu}_{~\mu} = & \frac{(2\pi)^4}{2N_c}\frac{S}{M^2} 
\int d^2\kperpone d^2\kperptwo\delta^2(\qvecperp-\kvecperpone-\kvecperptwo)
\\\label{eq:3.2}
&\times \sum_{i=1}^{N_f}e_i^2\left[
f_{i}(x_1,\kvecperpone,M) {f}_{\bar{i}}(x_2,\kvecperptwo,M)+(1\leftrightarrow 2)\right] 
{\rm Tr}\!\left[\kslash_1\gamma^\mu\,\kslash_2\gamma_\mu\right]
\end{align}
where $\kvecperpone,\kvecperptwo$ are quark transverse momenta, $x_{1,2}$ are their longitudinal momenta 
and $f_{i}, {f}_{\bar{i}}$ are transverse momentum dependent quark/antiquark distributions taken at the scale $Q=M$. 
\subsection{On-shell matrix element}
\label{DY_Kw_onshell_section}

The trace in (\ref{eq:3.2}) is the squared matrix element of the process $q(k_1)\qbar(k_2)\to \gamma^*$ in the lowest order.
For the on-shell matrix element, we use the quark/antiquark momenta in the collinear approximation
\be
k_1=x_1P_1\,,~~~~~~~~~~~~~~~~~k_2=x_2P_2\,.
\ee
what assures gauge invariance of the matrix elements.
In such a case
\be\label{eq:3.5bc}
{\rm Tr}\!\left[\kslash_1\gamma^\mu\,\kslash_2\gamma_\mu\right]=x_1x_2 {\rm Tr}\!\left[\Pslash_1\gamma^\mu\Pslash_2\gamma_\mu\right] = -4Sx_1x_2
\ee
and the DY cross section (\ref{eq:3.1xxx}) is given by 
\begin{align}\nonumber
\frac{d\sigma^{DY}}{dy_\gamma\,dM^2 \,d^2\qperp} = &
\frac{4\pi\alpha_{\rm em}^2}{3N_c\,M^4} \int d^2\kperpone d^2\kperptwo\,\delta^2(\qvecperp-\kvecperpone-\kvecperptwo)
\\\label{eq:3.68}
&\times 
\sum_{i=1}^{N_f} e_i^2\,x_1x_2\! \left[f_{i}(x_1,\kvecperpone, M)\, f_{\bar{i}}(x_2,\kvecperptwo,M)+(1\leftrightarrow 2)\right],
\end{align}
It is easy to check that after integrating (\ref{eq:3.68}) over $\qvecperp$, we find the leading order form of the Drell-Yan cross section 
with collinear PDFs given by eq.\,(\ref{eq:2.28aaa})
\be
\frac{d\sigma^{DY}}{dy_\gamma\,dM^2}= \frac{\sigma_0}{M^4}\sum_{i=1}^{N_f} e_i^2\,x_1x_2\!\left[ q_i(x_1,M^2) \bar{q}_{{i}}(x_2,M^2)+(1\leftrightarrow 2)\right],
\ee
where $\sigma_0 = {4\pi\alpha_{\rm em}^2}/{3N_c }$.
Inserting the delta function
\be\label{eq:5.13x}
\delta^2(\kvecperpone+\kvecperptwo -\qvecperp)=\int\frac{d^2b}{(2\pi)^2}\,\eto^{i(\kvecperpone+\kvecperptwo-\qvecperp)\cdot\bvec}
\ee
to eq.~(\ref{eq:3.68}), we find the DY cross section with the $b$-dependent parton distributions (\ref{Fourier_transf})
\begin{align}\label{eq:3.8a}
\frac{d\sigma^{DY}_{\rm on-shell} }{dy_\gamma\,dM^2 \,d^2\qperp} =
\frac{\sigma_0}{M^4}
\int \frac{d^2\bvec}{(2\pi)^2}\,\eto^{-i\qvecperp\cdot\bvec}\,
\sum_{i=1}^{N_f} e_i^2\,x_1x_2\!\left[\ftilde_{i}(x_1,\bvec, M)\, {\ftilde}_{\bar{i}}(x_2, \bvec,M)+(1\leftrightarrow 2)\right].
\end{align}
For the parton distributions which depend on $b=|\bvec|$, the angular integration with the help of relation (\ref{eq:2.30a}) gives
\begin{align} \label{DY_Kw_xs}
\frac{d\sigma^{DY}_{\rm on-shell} }{dy_\gamma\,dM^2 \,d\qperp^2} =
\frac{\sigma_0}{M^4}
\int_0^\infty \frac{bdb}{2}J_0(\qperp b)\sum_{i=1}^{N_f} e_i^2\,x_1x_2\!\left[\ftilde_{i}(x_1, b, M)\, {\ftilde}_{\ibar}(x_2, b,M)+(1\leftrightarrow 2)\right].
\end{align}
We will use this expression for the comparison with the DY data using the parton distributions which are solutions of the CCFM-K equations with the momentum fractions in the on-shell form
\be\label{eq:xfrac1}
x_{1,2}={\frac{M}{\sqrt{S}}}\,\eto^{\pm y_\gamma}\,.
\ee

\subsection{Off-shell matrix elements}
\label{DY_Kw_offshell_section}

In approach with the off-shell matrix, the trace (\ref{eq:3.5bc}) is replaced by
\be\label{eq:3.12}
{\rm Tr}\!\left[\kslash_1\gamma^\mu\,\kslash_2\gamma_\mu\right]~\to~{\rm Tr}\!\left[(x_1\Pslash_1)\Gamma^\mu(x_2\Pslash_2)\Gamma_\mu\right],
\ee
where $\Gamma^\mu$ is the Fadin-Sherman photon-quark vertex \cite{Fadin:1976nw,Lipatov:2000se} 
\be\label{FadSher_vert}
\Gamma^\mu=\Gamma^\mu(k_1,k_2)=\gamma^\mu-\frac{2\slashed{k}_1}{x_2S}P_1^\mu-\frac{2\slashed{k}_2}{x_1S}P_2^\mu.
\ee
and the quark/antiquark momenta $k_{1,2}$ take into account transverse components
\be\label{eq:3.11aa}
k_1=x_1P_1+\kperpone\,,~~~~~~~~~~~~~~~~~~~~~~~~k_2=x_2P_2+\kperptwo,
\ee
They are given by $\kperp_{i}=(0,\kvec_{T i},0)$ for $i=1,2$ while the momentum fractions $x_{i}$ are determined from the momentum conservation at the vertex, $q^2=(k_1+k_2)^2$,
which gives
\be\label{eq:3.18}
x_{1,2}={\frac{M_T}{\sqrt{S}}}\,\eto^{\pm y_\gamma}\,,~~~~~~~~M_T=\sqrt{M^2+\qperp^2}.
\ee
It is easy to check that the Fadin-Sherman vertex obeys the gauge invariance relation
\be
(k_1+k_2)_\mu\Gamma^\mu(k_1,k_2)=0.
\ee
Computing the trace (\ref{eq:3.12}), we obtain
\begin{align}\nonumber
{\rm Tr}\!\left[(x_1\Pslash_1)\Gamma^\mu(x_2\Pslash_2)\Gamma_\mu\right] &= -4Sx_1x_2\left(1-\frac{2\vec{k}_{1T}\cdot \vec{k}_{2T}}{M_T^2}\right) 
\\
&=-4Sx_1x_2\left(\frac{M^2+\kvecperpone^2+\kvecperptwo^2}{M_T^2}\right)
\end{align}
where we used momentum conservation at the photon vertex to write the last equality. Notice that because
of the transverse mass $M_\perp$ in the denominator, the off-shell kinematics takes into account the corrections in powers of $q^2_\perp/M^2$ to all orders.
With thes results, the cross section (\ref{eq:3.1xxx}) is given by
\begin{align}\nonumber
\frac{d\sigma^{DY}_{\rm off-shell}}{dy_\gamma\,dM^2 \,d^2\qperp} = &
\frac{4\pi\alpha_{\rm em}^2}{3N_c\,M^2M_T^2} \int d^2\kperpone d^2\kperptwo\,\delta^2(\qvecperp-\kvecperpone-\kvecperptwo)
\left(1+\frac{\kvecperpone^2+\kvecperptwo^2}{M^2}\right) 
\\\label{eq:3.68xxx}
&\times 
\sum_{i=1}^{N_f} e_i^2 x_1x_2\left[f_{i}(x_1,\kvecperpone, M)\, f_{\ibar}(x_2,\kvecperptwo,M)+(1\leftrightarrow 2)\right].
\end{align}
Inserting the delta function (\ref{eq:5.13x}) and performing the Fourier transformation, 
we obtain the following cross section with the parton distributions which depend on $b=|\bvec|$
\begin{align}\nonumber
&\frac{d\sigma^{DY}_{\rm off-shell}}{dy_\gamma\,dM^2 \,d\qperp^2} =
\frac{\sigma_0}{M^2M_\perp^2}
\int_0^\infty \frac{bdb}{2}\,J_0(\qperp b)\,\sum_{i=1}^{N_f} e_i^2\,x_1x_2
\, \bigg\{\ftilde_{i}(x_1,b,M) \ftilde_{\ibar}(x_2,b,M)\,+
\\\label{off-shell_CCFM-K_bspace}
&~~~~\,-\frac{1}{M^2}\left(\Delta_b\ftilde_{i}(x_1,b,M)\, \ftilde_{\ibar}(x_2,b,M) +
f_{i}(x_1,b,M)\, \Delta_b \ftilde_{\ibar}(x_2,b,M)\right)
+(1\leftrightarrow 2)
\bigg\}
\end{align}
where $\Delta_b$ is the radial part of the two-dimensional Laplacian 
\be
\Delta_b=\frac{\partial^2}{\partial b^2}+\frac{1}{b} \frac{\partial}{\partial b}\,.
\ee

By the comparison with the cross section (\ref{DY_Kw_xs}), we see that (\ref{off-shell_CCFM-K_bspace}) 
has different mass dependence, 
\be
d\sigma^{DY}_{\rm off-shell} \sim \frac{\sigma_0}{M^4(1+\qperp^2/M^2)}\,,~~~~~~~~{\rm vs.}\,~~~~~~~~
d\sigma^{DY}_{\rm on-shell} \sim \frac{\sigma_0}{M^4}\ .
\ee
It should be emphasized that the corrections $\qperp^2/M^2$ which are resummed to the factor $1/M_\perp^2$ in (\ref{off-shell_CCFM-K_bspace}) are entirely due to off-shellness of the matrix element. In the CSS approach such corrections, if large, signal the breaking of the CSS approximation. However, in the approach with transverse momentum dependent parton distributions (like the CCFM-K approach), they are naturally incorporated in the PDFs and off-shell matrix element. This is the main advantage of 
this method.
\begin{figure}[t]
\begin{center}
\includegraphics[width=0.6\textwidth]{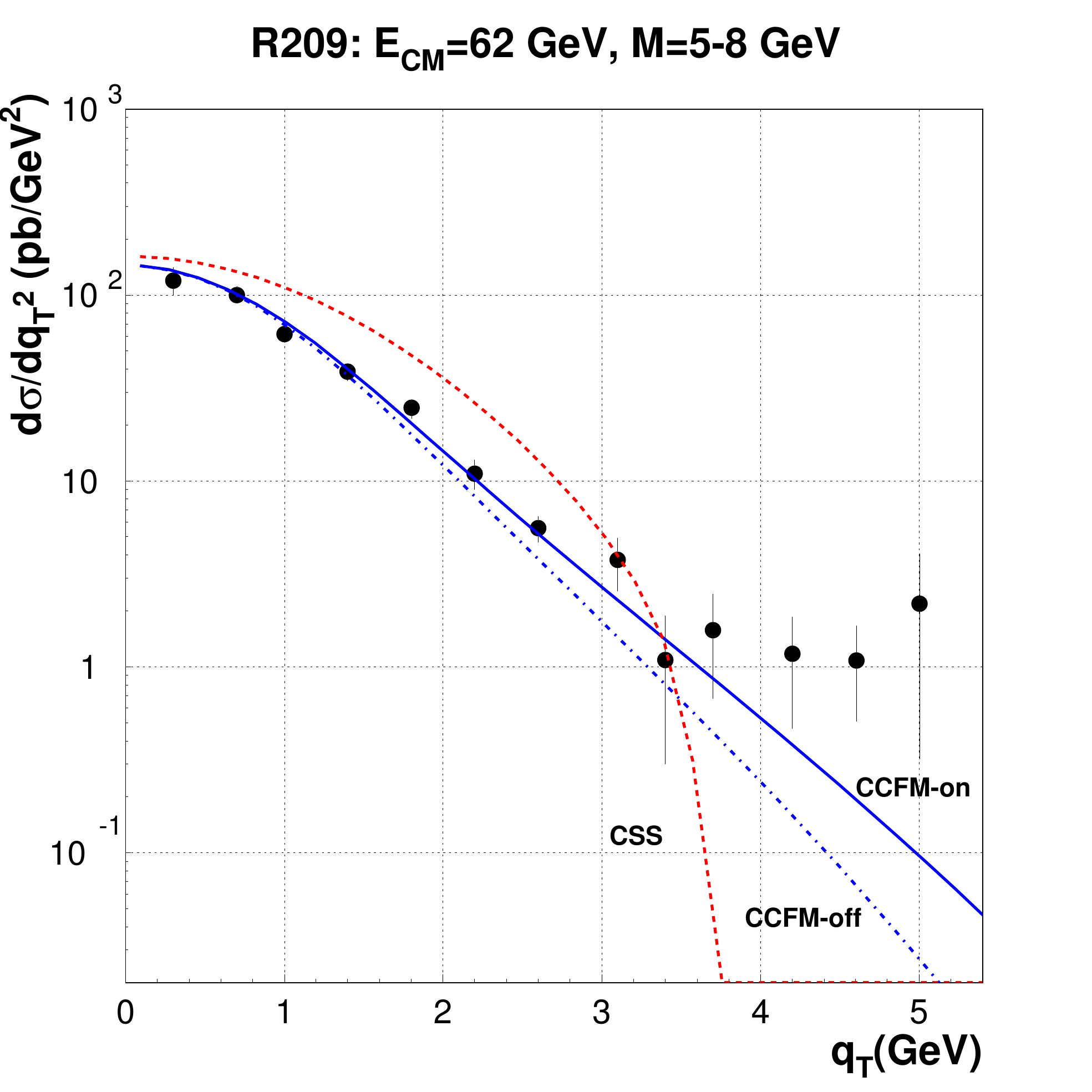}
\end{center}
\vskip -5mm
\caption{Transverse momentum dependence of the DY cross sections: data from proton-proton 
R209 experiment are compared with CCFM-K on-shell cross section (\ref{DY_Kw_xs}) 
(solid line), CCFM-K off-shell cross section (\ref{off-shell_CCFM-K_bspace}) 
(dash-dotted line) and CSS cross section (\ref{csxdyb1}) (dashed line).}
\label{fig:4}
\end{figure}

Numerical studies show that the contribution from the terms in the second line in (\ref{off-shell_CCFM-K_bspace}) is negligible. Therefore, for the same values of $x_1$ and $x_2$, the cross section $d\sigma^{DY}_{\rm off-shell}$ 
is suppressed by a factor $M^2/M_\perp^2$ in comparison to $d\sigma^{DY}_{\rm on-shell}$. 
In addition, in the off-shell case the PDFs are taken at larger values of $x_1$ and $x_2$, compare (\ref{eq:xfrac1}) and (\ref{eq:3.18}), which additionally suppresses $d\sigma^{DY}_{\rm off-shell}$ at large $q_\perp$.
This effect is clearly visible in Fig.\,\ref{fig:4} where we plot the CCFM-K predictions against
the Fermilab R209 data \cite{Antreasyan:1981eg}. The solid line corresponds to the on-shell cross section (\ref{DY_Kw_xs}) with the CCFM-K parton distributions evolved 
from the initial conditions (\ref{eq:4.10}) at $Q_0=1~{\rm GeV}$ with the MSTW08 LO PDFs \cite{Martin:2009iq} and the form factor (\ref{eq:2.37a}) with $b_0=2.7~{\rm GeV}^{-1}$. 
The dash-dotted line is obtained from the off-shell cross section (\ref{off-shell_CCFM-K_bspace}) with the same parton distributions. 
We also show, for general orientation, the prediction from the CSS formula (\ref{csxdyb1}) (red dashed line) which is discussed in detail in the next section.

\subsection{CSS approach}
\label{DY:CSS}

The CSS approach to the DY process has a long history which starts with the pioneering work
\cite{Collins:1984kg}. In this approach, collinearly colliding quarks emit 
gluons with a total transverse momentum $\qperp$ which is balanced
by the transverse momentum of the DY boson. The soft and collinear divergences for $\qperp\to 0$ in real emission
are not fully cancelled by virtual corrections and manifest themselves by the presence
of large logarithms, $\log(M/\qperp)$, which are resummed in the CSS approach. This leads for example to
evolution equations with two scales in the NNLL approximation. The recent state-of-art analyses of the
DY data with the CSS approach up to the order N$^3$LL were done in \cite{Bacchetta:2019sam,Scimemi:2019cmh}.

Such a precision in the CSS approach is beyond the scope of our present analysis since we do not aim at a comprehensive description of the data using this approach. It only serves for the comparison with the results of the CCFM-K approach in 
which $\qperp$ of the DY boson is the sum of intrinsic transverse momenta of colliding partons, see the formulae (\ref{eq:3.1xxx}) and (\ref{eq:3.2}). For this reason, we also do not consider the so-called ``Y term", which was proposed in \cite{Collins:1984kg} to match the CSS formula with the fixed-order results. Thus, we will use the CSS formulae in the NLL approximation with one scale evolution.
Nevertheless, all the problems which are encountered in the description of the DY data in this approximation are still present in the analyses done with higher order approximations.

The DY cross section in the CSS approach up to the NLL order is given by
\be
{d \sigma^{DY}_{CSS} \over d y_\gamma dM^2 d q_T^2}
= \frac{\sigma_0}{M^2}
\int_0^\infty \frac{bdb}{2}J_0(q_T b)
\sum_{i=1}^{N_f} e_i^2\left\{W_{i\ibar}(x_1,x_2,b,Q)+ (1\leftrightarrow 2)\right\},~~~~
\label{csxdyb1}
\ee
where $Q=M$ and $x_{1,2}={M}\eto^{\pm y_\gamma}/{\sqrt{S}}$. The parton luminosities $W_{i\ibar}$ in the above read
\be\label{eq:3.21aa}
W_{i\ibar}(x_1,x_2,b,Q)=
f_{i}^\prime(x_1,c/b_{*})\,f_{\ibar}^\prime(x_2,c/b_{*})\,e^{2S(b_*,Q)} \, W_{NP}(x_1, x_2,b,Q),
\label{swij}
\ee
where $f_{i/\ibar}^\prime$ are effective quark/antiquark distributions 
\be\label{eq:4.3a}
f^\prime_{i/\ibar}(x,\mu) =\int_x^1\frac{dz}{z} \left\{C_q({x}/{z},\alpha_s(\mu)))\,q_{i/\ibar}(z,\mu)
+C_g({x}/{z},\alpha_s(\mu))\,g(z,\mu)\right\}
\ee
with the $\overline{\rm MS}$ NLO collinear PDFs 
$q_{i/\ibar}(z)$ and $g(z)$ and the coefficient functions 
\begin{align}\nonumber
C_{q}(z,\alpha_s) &=\delta(1-z)+\frac{\alpha_s}{2\pi}C_F\left[1-z+\left(\frac{\pi^2}{2}-4\right)\delta(1-z)\right\},
\\
C_{g}(z,\alpha_s) &= \frac{\alpha_s}{2\pi}T_R\left[2z(1-z)\right].
\end{align}
The scale $\mu$ in (\ref{eq:4.3a}) is given by $\mu=c/b_*$ with $c=2\eto^{-\gamma_E}\approx 1.12$ and 
\be\label{eq:3.25zz}
b_*=\frac{b}{\sqrt{1+{b^2}/{b_{\rm max}^2}}}.
\ee
TIn this way, $b_*$ interpolates between $b_*=0$ and $b_*=b_{\rm max}$ for $b\to \infty$ such that the scale
\be
\mu\in [c/b_{\rm max},\infty)
\ee
Thus, choosing $b_{\rm max}=c/Q_0$, where $Q_0$ is an initial
scale for the DGLAP evolution, we ensure that the collinear PDFs are always defined during the integration over $b$ in (\ref{csxdyb1}). In our presentation, we use the MSTW08 NLO PDFs \cite{Martin:2009iq} and choose $Q_0=1~{\rm GeV}$.

The power $S$ in the exponent in (\ref{eq:3.21aa}) is given by
\be\label{eq:3.21xx}
S(b,Q)\,=\,-\int_{c^2/b^2}^{Q^2}{d q^2 \over q^2}\left[A_q(\alpha_s(q^2))\ln\left({Q^2\over q^2}\right)+ B_q(\alpha_s(q^2))\right]
\label{sibq}
\ee
where the coefficients $A_q$ and $B_q$ are defined by the general perturbative expansion 
\be
A_q(\alpha_s)=\sum_{n=1}^{\infty}\left({\alpha_s\over 2\pi}\right)^nA_q^{(n)}
\,,~~~~~~~~~~~
B_q(\alpha_s)=\sum_{n=1}^{\infty}\left({\alpha_s\over 2\pi}\right)^nB_q^{(n)}\,.
\label{abn}
\ee
Introducing $B=\ln(Q^2b^2/c^2)$ and $L=L(Q)=\ln(Q^2/\Lambda^2)$, 
the LL approximation is defined by the terms proportional to $B(B/L)^n$ while in the NLL approximation terms proportional 
to $(B/L)^n$ are added. Thus, in he NLL approximation which we consider, the coefficients are given by
\cite{Kodaira:1981nh,Kodaira:1982az,Catani:1988vd}
\be
A_{q}^{(1)}=C_F\,,~~~~~~~~~~~~~A_{q}^{(2)}= C_FK \,,~~~~~~~~~~~~~B_{q}^{(1)}=-{\frac{3}{2}}C_F
\label{coeff_NLL}
\ee
and $K=C_A(\textstyle{\frac{67}{18}}-\textstyle{\frac{1}{6}}\pi^2)-\textstyle{\frac{10}{9}}T_RN_f$. 
By the comparison of the power $S$ given by (\ref{eq:3.21xx}) with that in (\ref{eq:2.38ab}), 
we see that the CCFM-K equations only partially resum the next-to-leading logarithms
since the term proportional to $A_q^{(2)}$, which is formally of the NLL accuracy, is missing in (\ref{eq:2.38ab}).
It can be obtained, however, from the CCFM-K equations with the higher order splitting functions.
Using the two-loop running coupling constant
\be
\alpha_s(q^2)=\frac{1}{\beta_0 L(q)} -\frac{\beta_1}{\beta_0^3}\frac{\ln L(q)}{L^2(q)}
\ee
where $\beta_0=(11C_A-4T_RN_f)/12\pi$ and $\beta_1=(153-19N_f)/24\pi^2$, and performing the integration in (\ref{eq:3.21xx}), 
one obtains the final NLL form of $S$ \cite{Catani:1988vd}, which we use in our presentation
\begin{align}\nonumber
S = &\frac{A_q^{(1)}}{\pi\beta_0}\left[L\ln\!\left(1-\frac{B}{L}\right) +B\right] - \frac{A_q^{(2)}}{\pi^2\beta_0^2}\left[ \ln\!\left(1-\frac{B}{L}\right)+\frac{B}{L-B}\right] 
+\frac{B_q^{(1)}}{\pi\beta_0} \ln\!\left(1-\frac{B}{L}\right)
\\\label{eq:3.25aa}
&+ \frac{A_q^{(1)}\beta_1}{\pi\beta_0^3}\left[ \frac{B}{L-B} (1+\ln L)+\left(\frac{L}{L-B}+\ln L\right) \ln\!\left(1-\frac{B}{L}\right) +\frac{1}{2} \ln^2\!\left(1-\frac{B}{L}\right)
\right].
\end{align}

The factor $W_{NP}=\exp\{-S_{NP}\}$ in (\ref{swij}) describes the non-perturbative contribution \cite{Qiu:2000hf,Berger:2002ut,Kulesza:2003wi}. 
In our presentation, we use the form factor from the BLNY fit \cite{Landry:2002ix} to the DY data:
\be
S_{NP} = [a_1+a_2\ln({Q}/{Q_1}) + a_3\ln(100\, x_1x_2)]\, b^2,
\label{BLNY_form}
\ee
where 
\be
a_1=0.21~{\rm GeV}^{2}\,,~~~~~~~a_2=0.68~{\rm GeV}^{2}\,,~~~~~~~a_3=-0.1~{\rm GeV}^{2}\,,~~~~~~~Q_1=3.2~{\rm GeV}.
\ee

\section{Comparison to data}
\label{sect_comp_data}

For the comparison with the low mass DY data, we use the CCFM-K approach with both on-shell and off-shell matrix elements (see sections \ref{DY_Kw_onshell_section} and \ref{DY_Kw_offshell_section}). In this approach, we only have one free parameter, $b_0$ in the non-perturbative form factor (\ref{eq:2.37a}), which we somewhat optimized to the value $b_0=2.7~{\rm GeV}^{-1}$. For the initial PDFs we use the MSTW08 LO PDF set \cite{Martin:2009iq}. 

We also show the CSS results at the NLL accuracy with the BLNY form factor (\ref{BLNY_form}) and the MSTW08 NLO PDF set \cite{Martin:2009iq} (see the previous section). We use this more refined form factor and PDF sets (compared to those of CCFM-K) in order to reach better description of data within the CSS formalism at the NLL accuracy. 
The results depend to some extent on the value of the parameter $b_{\rm max}=c/Q_0$ in eq.~(\ref{eq:3.25zz}) but not such that the general conclusions concerning the CSS description should be changed. For example, using $Q_0=2\,{\rm GeV}$ makes the curves
 stronger suppressed for large $q_T$.

Any attempt to have exactly the same set of parameters for both the CSS and CCFM-K approaches leads to significant deterioration
of the agreement with the data in one or the other description. This is not a surprise since the CSS and CCFM-K approaches have different starting points (collinear versus $k_T$ -- factorization) and are derived in different approximations (NLL versus LL).
Therefore, they have to be optimized with respect to the DY data description separately. To check the impact of the choices we made, we produced the results for NLL CSS with the Gaussian form factor (\ref{eq:2.37a}) (with properly chosen $b_0$) and MSTW08 LO PDF. It turns out that in such a case, the description of data at small $q_T$ is worse than for the one with the NLO PDFs and the non-perturbative contribution (\ref{BLNY_form}). Moreover, as expected, the rapid fall of the CSS curves at high $q_T$ is still present 
(see also \cite{Bacchetta:2019tcu} for detailed discussion of difficulties with matching the CSS approach to fixed order calculation and description of data at $q_T\sim M$).

\subsection{DY from fixed target experiments}

We start with the data from the fixed target experiments E288 \cite{Ito:1980ev}, E605 \cite{Moreno:1990sf}, E866 \cite{Webb:2003bj} and E722 \cite{McGaughey:1994dx}. The cross section $E d^3 \sigma/d^3 q$ measured in these experiments is related to (\ref{DY_Kw_xs}) and (\ref{csxdyb1}) as follows
\be
E \frac{d^3 \sigma}{d^3 q} =\frac{2 M\Delta M}{\pi} \frac{d\sigma^{DY}}{dy_\gamma \,dM^2 \,d\qperp^2}
\label{Jacobian_fixTarg}
\ee
where $\Delta M$ is the bin size of the DY pair mass distribution. In addition, the E605, E866 and E722 experiments also measured the cross section in bins of the Feynman variable
\be
x_F \equiv \frac{2q^3}{\sqrt{S}}= \frac{2\sqrt{M^2+q^2_T}}{\sqrt{S}} \sinh{y_\gamma},
\ee 
where the r.h.s. gives the relation between $x_F$ and the DY photon rapidity $y_\gamma$. The energies of the proton projectile were equal to: 200, 300 and 400 GeV (at E288), and 800 GeV (at E605, E866 and E772). These translate into the center of mass energies $\sqrt{S}= 19.4,\ 23.8, \ 27.4$ and 38.8~GeV, respectively. The experiments differ by the targets used: E288 used Cu or Pt, E605 used Cu, E866 used H or D and E772 used $^2$H. All the cross sections were normalized by the number of nucleons in the target nucleus. In what follows, we neglect nuclear effects of the targets and compare unmodified CCFM-K and CCS approaches with such data.

In Fig.\,\ref{fig:e288} we show the data from the E288 experiment \cite{Ito:1980ev} for three values of energies and rapidities. At each panel, the transverse momentum dependence of the DY cross section is shown with fixed mass $M$ and rapidity $y$ (or $x_F$)\footnote{In most cases we use the centers of the bins in $M$ and $x_F$, after checks that the experimental bin sizes are not very important.}. The mass range equals $M=4.5-12.5$~GeV. 
In Fig.\,\ref{fig:e605_e866} we show the data from the E605 \cite{Moreno:1990sf} and E866 experiments \cite{Webb:2003bj} for $\sqrt{S}=38.8$~GeV,
$x_F=0.1$ and the mass range $M=4.7-15.5$~GeV. In Fig.\,\ref{fig:e866} we show the data from the E866 experiment for $\sqrt{S}=38.8$~GeV, three values of $x_F$ and the mass range $M=4.7-14.85$~GeV. Finally, in Fig.\,\ref{fig:E772} we show the data from the E772 experiment \cite{McGaughey:1994dx} 
for $\sqrt{S}=38.8$~GeV, $0.1<x_F<0.3$ and the mass range $M=5.5-14.5$~GeV. 

The data in Figs.\,\ref{fig:e288}--\ref{fig:E772} are compared to theoretical curves: CCFM-K on-shell (blue solid curves), CCFM-K off-shell (blue dashed-dotted curves) and CSS (red dashed curves). Comparing CCFM-K to CSS we see that at small $q_T$ the CSS resummation predicts higher cross-section than CCFM-K and better agrees with the E288 and E605 data. This comes from the fact that the parameters of the non-perturbative form factor (\ref{BLNY_form})
were fitted in \cite{Landry:2002ix} to the data while in the CCFM-K approach we fixed just one free parameter $b_0$ in the form factor (\ref{eq:2.37a}) to $b_0=2.7\,{\rm GeV}^{-2}$. 

The motivation for that was to show the potential of the CCFM-K approach to describe the large $q_T$ data
without going into details of fitting the parameter $b_0$, as in the CSS approach. Thus, at larger $q_T$ (2-3 GeV, depending on $M$ and $x_F$), the CSS curves drop rapidly as we do not match them to the fixed order calculation by adding the ``Y term".
On the other hand, the CCFM-K curves describe the data reasonably well. Note also that for 
$M \sim 9\,{\rm GeV}$, the data from E288 are significantly above the theoretical predictions which is related to the production of $\Upsilon$ meson, not considered in our calculations.

The E866 and E772 data seems to be systematically above theoretical predictions at small $q_T$, except for a few values of $M$ and $x_F$. As before, CCFM-K provides good description of the data at large $q_T$. In general, one sees better description of data for higher DY masses. 

Comparing the CCFM-K on-shell and off-shell approaches one sees that the former approach agrees better with data as providing slower drop with $q_T$. The difference is larger for $q_T\sim M$, as one should expect, see discussion at the end of section \ref{DY_Kw_offshell_section}.

\subsection{DY in proton-proton collisions}

We also consider the data from two experiments measuring the DY production in proton-proton collisions at moderate energies: R209 \cite{Antreasyan:1981eg} with $\sqrt{S}= 62$ GeV and PHENIX \cite{Aidala:2018ajl} with $\sqrt{S}= 200$ GeV. For R209 we apply a change of variables,
\be
\frac{d \sigma}{d^2 q_T}= \int\limits_{5-8\textrm{ GeV}} d M \frac{M\sqrt{S}}{\sqrt{M^2 + q_T^2}} \frac{d\sigma^{DY}}{dy\,dM^2 \,dq_T^2},
\ee 
whereas PHENIX is using the cross section $E d^3 \sigma/d^3 q$ given by (\ref{Jacobian_fixTarg}).

The theoretical results were compared with the R209 data in Fig.\,\ref{fig:4}. CCFM-K provides very good description of data up to $q_T\sim 4$~GeV and slightly underestimate cross-section for higher $q_T$ while CSS overestimates the cross section at small $q_T$ and decreases rapidly at high values. For the PHENIX data shown in Fig\,\ref{fig:PHENIX}, CCFM-K gives a better description than CSS, which overestimates the data at moderate values of $q_T$. We note that as for fixed target experiments, the on-shell CCFM-K better describes the data then the off-shell approach.
\section{Summary}
\label{sect_summary}

Using the CCFM-K parton distributions and the partonic cross-section with on-shell and off-shell matrix element, we analyze the transverse momentum spectra of the DY dileptons from all available low mass data.
The overall description of these data is quite good, given the simplicity of the non-perturbative Gaussian input (\ref{eq:2.37a}) with only one free parameter $b_0$, being the Gaussian width. We have chosen optimized value of this parameter for all experiments to show the potential of the CCFM-K approach in the description of the data for
large dilepton transverse momentum $q_T$. 

However, for small $q_T$ we found less successful description, especially in low mass bins. 
This calls for an approach with more non-perturbative parameters in initial conditions for the CCFM-K evolution akin to the BLNY fit \cite{Landry:2002ix} of the non-perturbative form factor (\ref{BLNY_form}) in the CSS
approach. This is justified since the CCFM-K evolution includes elements of the CSS resummation. In this sense our paper should be treated as a step towards unified description of the low mass DY data, where the CSS approach matched to the fixed order calculation experiences some troubles \cite{Bacchetta:2019tcu}.

One should also note that the presented analysis is based on the LO matrix elements and 
the CCFM-K evolution equations in the single loop approximation. For these reasons, we decided to postpone
the analysis with more complicated non-perturbative input to future studies with the NLO matrix elements and 
evolution equations. The first attempt in this direction was done recently in \cite{Martinez:2020fzs} using the Parton Branching method.
Finally, it is important to stress that the future analysis should also include the Tevatron and LHC data on the weak bosons production.

\section*{Acknowledgments}
We thank Hannes Jung, Leszek Motyka and Anna Sta\'sto for useful discussions. This work was supported by the National Science Center, Poland, Grants Nos.~2015/17/B/ST2/01838, 2017/27/B/ST2/02755, 
2019/33/B/ST2/02588 and 2019/32/C/ST2/00202.

\appendix

\section{Relation of CCFM-K to CSS}
\label{app:A}

Following the method presented in \cite{Gawron:2003np}, we show the CCFM-K resummation contains the soft gluon resummation of Collins, Soper and Sterman (CSS) \cite{Collins:1984kg}. In order to simplify the notation, we apply the Mellin transform
\be
\fbar(n,b,Q)=\int_0^1 dx\,x^{n} \ftilde(x,b,Q)
\ee
to both sides of eqs.~(\ref{eq:4.8a}) to obtain
\begin{align}\nonumber
\frac{\partial \fbar_i(n,b,Q)}{\partial \ln Q^2}&= \Pbar_{qq}(n,b,Q)\,\fbar_i(n,b,Q)+ \Pbar_{qg}(n,b,Q)\,\fbar_g (n,b,Q),
\\\label{eq:2.36a}
\frac{\partial \fbar_g(n,b,Q)}{\partial \ln Q^2}&= \Pbar_{gq}(n,b,Q)\, \fbar_S(n,b,Q) + \Pbar_{gg}(n,b,Q)\, \fbar_g(n,b,Q),
\end{align} 
where $\fbar_S=\sum_{i=1}^{2N_f} \fbar_i$ is the singlet quark distribution and the Mellin moments of the splitting functions read
\begin{align}\nonumber
\Pbar_{qq}(n,b,Q) &= \frac{\alpha_s(Q^2)}{2\pi} \int_0^1dz\,P_{qq}(z)\left\{z^n J_0((1-z)Qb)-1\right\},
\\\nonumber
\Pbar_{qg} (n,b,Q)&= \frac{\alpha_s(Q^2)}{2\pi} \int_0^1dz\,P_{qg}(z)\,z^n J_0((1-z)Qb),
\\\nonumber
\Pbar_{gq}(n,b,Q)&= \frac{\alpha_s(Q^2)}{2\pi}\int_0^1dz\,P_{gq}(z) \,z^n J_0((1-z)Qb),
\\
\Pbar_{gg} (n,b,Q)&= \frac{\alpha_s(Q^2)}{2\pi} \int_0^1dz \left\{P_{gg}(z)\, z^n J_0((1-z)Qb) -z[P_{gg}(z)+2N_fP_{qg}(z)]\right\} .
\end{align}
Notice that for $b=0$ in (\ref{eq:2.36a}) we obtain the DGLAP evolution equations. This fact motivates the following decomposition of the diagonal splitting functions
\begin{align}\nonumber
\Pbar_{qq}(n,b,Q) &= \Pbar_{qq1}(n,b,Q) + \Pbar_{qq2}(b,Q),
\\
\Pbar_{gg}(n,b,Q) & = \Pbar_{gg1}(n,b,Q) + \Pbar_{gg2}(b,Q),
\end{align}
where
\begin{align}\label{eq:a5}
\Pbar_{qq1}(n,b,Q) &= \frac{\alpha_s(Q^2)}{2\pi} \int_0^1dz\,P_{qq}(z)(z^n-1) J_0((1-z)Qb),
\\\label{eq:a6}
\Pbar_{qq2}(b,Q) &= \frac{\alpha_s(Q^2)}{2\pi} \int_0^1dz\,P_{qq}(z)\left\{J_0((1-z)Qb)-1\right\},
\\\label{eq:a7}
\Pbar_{gg1}(n,b,Q) &= \frac{\alpha_s(Q^2)}{2\pi} \int_0^1dz \left\{ z^n P_{gg}(z)- z[P_{gg}(z)+2N_fP_{qg}(z)]\right\} J_0((1-z)Qb),
\\\label{eq:a8}
\Pbar_{gg2}(b,Q) &= \frac{\alpha_s(Q^2)}{2\pi} \int_0^1dz \,z[P_{gg}(z) + 2N_f P_{qg}(z)]\left\{J_0((1-z)Qb) -1\right\}.
\end{align} 
Thus, for $b=0$, $\Pbar_{qq2}=\Pbar_{gg2}=0$, and $\Pbar_{qq1}$ and $\Pbar_{gg1}$ become the ordinary Altarelli-Parisi splitting functions.
This is why we write (\ref{eq:2.36a}) in the form
\begin{align}\nonumber
\frac{\partial \fbar_i}{\partial \ln Q^2}- \Pbar_{qq2}\, \fbar_i&= \Pbar_{qq1}\,\fbar_i + \Pbar_{qg}\,\fbar_g,
\\
\label{eq:2.39a}
\frac{\partial \fbar_g}{\partial \ln Q^2}- \Pbar_{gg2}\, \fbar_g&= \Pbar_{gq}\, \fbar_S + \Pbar_{gg1}\, \fbar_g,
\end{align}
where for simplicity we suppressed the arguments. We look for the solutions in the form 
\be\label{eq:2.40a}
{\fbar}_i(n,b,Q)=\eto^{S_q(b,Q)}\bar{\fbar}_i(n,b,Q)\,,~~~~~~~~~
{\fbar}_g(n,b,Q)=\eto^{S_g(b,Q)}\bar{\fbar}_g(n,b,Q),
\ee
where
\be\label{eq:2.44a}
S_q(b,Q)= \int_{Q_0^2}^{Q^2}\frac{dq^2}{q^2} \Pbar_{qq2}(b,q)\,,~~~~~~~~~
S_g(b,Q)= \int_{Q_0^2}^{Q^2}\frac{dq^2}{q^2} \Pbar_{gg2}(b,q).
\ee
Inserting (\ref{eq:2.40a}) to (\ref{eq:2.39a}), we find
\begin{align}\nonumber
\frac{\partial \bar{\fbar}_i}{\partial \ln Q^2} &= \Pbar_{qq1}\,\fbarbar_i + \left(\eto^{S_g-S_q}\Pbar_{qg}\right) \fbarbar_g,
\\
\label{eq:2.45a}
\frac{\partial \bar{\fbar}_g}{\partial \ln Q^2} &= \left(\eto^{S_q-S_g}\Pbar_{gq}\right) \fbarbar_S +\Pbar_{gg1}\, \fbarbar_g.
\end{align}
We are interested in the approximate solution when $b$ obeys relation (\ref{eq:2.39ab}), i.e. for
\be
Q_0\ll 1/b \ll Q\,.
\ee
In such a case, the powers of large logarithms, $\ln(Qb)$ and $\ln(1/Q_0 b)$, organize the calculations. 
We will show that in such an approximation the solution is given by the Mellin moments of the CSS formulas (\ref{eq:2.38ab})
\begin{align}\label{eq:2.47a}
\fbar_i(n,b,Q) &=\exp\bigg\{-\int_{1/b^2}^{Q^2}\frac{dq^2}{q^2}\frac{\alpha_s(q^2)}{2\pi}\left[A_{q}^{(1)}\ln(q^2b^2)+B_{q}^{(1)}\right]\bigg\}\,\bar{q}_i(n,1/b),
\\\label{eq:2.48a}
\fbar_g(n,b,Q) &=\exp\bigg\{-\int_{1/b^2}^{Q^2}\frac{dq^2}{q^2}\frac{\alpha_s(q^2)}{2\pi}\left[A_{g}^{(1)}\ln(q^2b^2)+B_{g}^{(1)}\right]\bigg\}\,\bar{g}(n,1/b)
\end{align}
where 
the coefficients
\be\label{eq:a16}
A_{q}^{(1)}=C_F\,,~~~~~~B_{q}^{(1)}=-\frac{3}{2}C_F\,,~~~~~~ A_{g}^{(1)}=C_A\,,~~~~~~B_{g}^{(1)}=\frac{2}{3}T_RN_f-\frac{11}{6} C_A.~~~
\ee
\paragraph{Proof.} 
The dominant contribution to the integrals with the Bessel function $J_0(u)$ comes from the region $u<1$, therefore, we use the following approximations
\be\label{eq:2.50a}
J_0(u)\approx \theta(c-u)\,,~~~~~~~~~~~1- J_0(u) \approx \theta(u-c),
\ee
where $\theta$ is the Heaviside step function and $c\sim 1$. The precise value of this parameter is important for numerical studies but for simplicity of 
this analysis we set $c=1$. Thus, the quark exponent $S_q$ in (\ref{eq:2.44a}) with $\Pbar_{qq2}$ given by (\ref{eq:a6}) is given by
\be
S_q(b,Q) = -\int_{Q_0^2}^{Q^2}\frac{dq^2}{q^2} \frac{\alpha_s(q^2)}{2\pi}\int_0^1 dz\,P_{qq}(z)\,\theta[(1-z)q b-1].
\ee
For $qb<1$, the argument of the theta function is negative and $S_q=0$. Thus, we have
\be\label{eq:2.55aa}
S_q(b,Q) = - \int_{Q_0^2}^{Q^2}\frac{dq^2}{q^2} \frac{\alpha_s(q^2)}{2\pi} \theta(q -1/b)\int_0^{1} dz\,P_{qq}(z)\,\theta(1-z-1/qb).
\ee
From the first theta function $q>1/b \gg Q_0$ which sets the lower integration limit to $1/b$. In this way, we avoid resummation of large logarithms 
$\ln(1/Q_0b)$ which are shifted to the functions $\bar{\fbar}_{i,g}$ in (\ref{eq:2.40a}) and need to be resummed separately. 
Since $1/b<Q$ in our approximation, the upper integration limit $Q^2$ in (\ref{eq:2.55aa}) is not affected by the first theta function.
Writing $P_{qq}$ in the form
\be
P_{qq}(z)=C_F\left(\frac{2}{1-z}-(1+z)\right),
\ee
we find the result which agrees with the exponent in (\ref{eq:2.47a})
\begin{align}\nonumber
S_q(b,Q) &= - \int_{1/b^2}^{Q^2}\frac{dq^2}{q^2} \frac{\alpha_s(q^2)}{2\pi} \int_0^{1-1/q b} dz\,C_F\left\{\frac{2}{1-z}-(1+z)\right\}
\\\label{eq:2.52a}
&\approx -\int^{Q^2}_{1/b^2}\frac{dq^2}{q^2} \frac{\alpha_s(q^2)}{2\pi} \left\{ C_F \ln(q^2 b^2) -\frac{3}{2}C_F\right\},
\end{align}
where in the last line we neglected terms with subleading powers of logs $\ln(Qb)$ after the integration. 
A similar calculation for $S_g$ in (\ref{eq:2.44a}) leads to the exponent (\ref{eq:2.48a})
\be
S_g(b,q) \approx -\int^{Q^2}_{1/b^2}\frac{dq^2}{q^2} \frac{\alpha_s(q^2)}{2\pi} \left\{ C_A \ln(q^2 b^2)+\frac{2}{3}T_RN_f -\frac{11}{6}C_A \right\}.
\label{Sudakov_Kw_gg}
\ee
The large logs $\ln(1/Q_0b)$ are resummed using the DGLAP evolution equations. To show this, we write (\ref{eq:2.45a}) in the integral form
\begin{align}
\fbarbar_i(n,b,Q) = \fbarbar_i(n,b,Q_0) &+ \int_{Q_0^2}^{Q^2} \frac{d\mu^2}{\mu^2}\Pbar_{qq1}(n,b,\mu)\,\fbarbar_i(n,b,\mu)~ +
\nonumber \\
&+ \int_{Q_0^2}^{Q^2} \frac{d\mu^2}{\mu^2}\eto^{S_g(b,\mu)-S_q(b,\mu)} \Pbar_{qg}(n,b,\mu)\,\fbarbar_g(n,b,\mu)\,.
\label{eq:2.53a}
\end{align}
The first integral on the r.h.s. is given by
\be
I_1= \int_{Q_0^2}^{Q^2} \frac{d\mu^2}{\mu^2} \frac{\alpha_s(\mu^2)}{2\pi}\left( \int_0^1dz\,P_{qq}(z)(z^n-1) J_0((1-z)\mu b)\right)\fbarbar_i(n,b,\mu)
\label{first_term_of_fbarbar}
\ee
where we used (\ref{eq:a5}). Approximating 
\be\label{eq:2.61aa}
J_0((1-z)\mu b)\approx \theta(1-(1-z)\mu b)\,,
\ee
we notice that in order to get the leading logs $\ln(1/Q_0b)$ we have to assume that $\mu b< 1$. 
In such a case, the integration over $z$ in (\ref{first_term_of_fbarbar}) is not constrained but
the integration over $\mu$ is limited to
$\mu<1/b\ll Q$. In this way
\be\label{eq:2.55a}
I_1\approx \int_{Q_0^2}^{1/b^2} \frac{d\mu^2}{\mu^2}\frac{\alpha_s(\mu^2)}{2\pi} \left(\int_0^1dz\,(z^n-1)P_{qq}(z)\right)\fbarbar_i(n,b,\mu)\,.
\ee
Applying the same approximations to second integral in (\ref{eq:2.53a}), we obtain 
\be\label{eq:2.56a}
I_2\approx \int_{Q_0^2}^{1/b^2} \frac{d\mu^2}{\mu^2}\frac{\alpha_s(\mu^2)}{2\pi}\,\eto^{S_q(b,\mu)-S_g(b,\mu)}\left(\int_0^1dzz^nP_{qg}(z)\right)\fbarbar_g(n,b,\mu) .
\ee
From (\ref{eq:2.52a}), in the integration region, $\mu <1/b$, we have $S_q(b,\mu)=S_g(b,\mu)=0$. Therefore, we can set the exponent in $I_2$ equal one and 
(\ref{eq:2.53a}) reads
\begin{align}\label{eq:2.65aa}
\fbarbar_i(n,b,1/b) = \fbarbar_i(n,b,Q_0) &+ \int_{Q_0^2}^{1/b^2} \frac{d\mu^2}{\mu^2}\frac{\alpha_s(\mu^2)}{2\pi} \left(\int_0^1dz\,P_{qq}(z)(z^n-1)\right) \,\fbarbar_i(n,b,\mu) ~+\nonumber \\
&+ \int_{Q_0^2}^{1/b^2} \frac{d\mu^2}{\mu^2}\frac{\alpha_s(\mu^2)}{2\pi}\, \left( \int_0^1dzz^nP_{qg}(z)\right) \,\fbarbar_g(n,b,\mu)\,.~~~~~~~
\end{align}
These are the DGLAP equation for the moments of quark the distributions evolved to the scale $Q=1/b$. The analogous considerations for the gluon distributions lead to the gluon counterpart of the DGLAP equations. It can easily be checked that for $\mu >1/b$, the theta function (\ref{eq:2.61aa}) imposes the condition $z>1-1/(\mu b)$ leads to subleading logarithms which we neglect. 

To make a connection with the collinear PDFs, we note that for the values of $b$ which we consider, 
$b\ll 1/Q_0\sim 1~{\rm GeV}^{-1}$, to good approximation $b\approx 0$ in the functions $\fbarbar_{i,g}$ in (\ref{eq:2.65aa}). 
Thus, choosing the initial conditions equal to the Mellin moments of the collinear PDFs, 
\be
\fbarbar_i(n,b\approx 0,Q_0) = \qbar_i(n,Q_0)\,,~~~~~~~~~~~~~~~\fbarbar_g(n,b\approx 0,Q_0) =\bar{g}(n,Q_0)\,,
\ee
we find the collinear PDFs at the scale $1/b$
\be
\label{fbarbar_rel_to_PDF}
\fbarbar_i(n,b\approx 0,1/b) = \qbar_i(n,1/b)\,,~~~~~~~~~~~~~~~\fbarbar_g(n,b\approx 0,1/b) =\bar{g}(n,1/b).
\ee 
This concludes the proof that (\ref{eq:2.47a}) and (\ref{eq:2.48a}) are the approximate solutions to the CCFM-K equations.
As a final remark, the parameter $c\ne 1$ in (\ref{eq:2.50a}) leads to the replacement
$1/b\to c/b$ in all the formulae above.

\begin{figure}[h]
\vskip -15mm
\begin{center}
\includegraphics[width=0.7\textwidth]{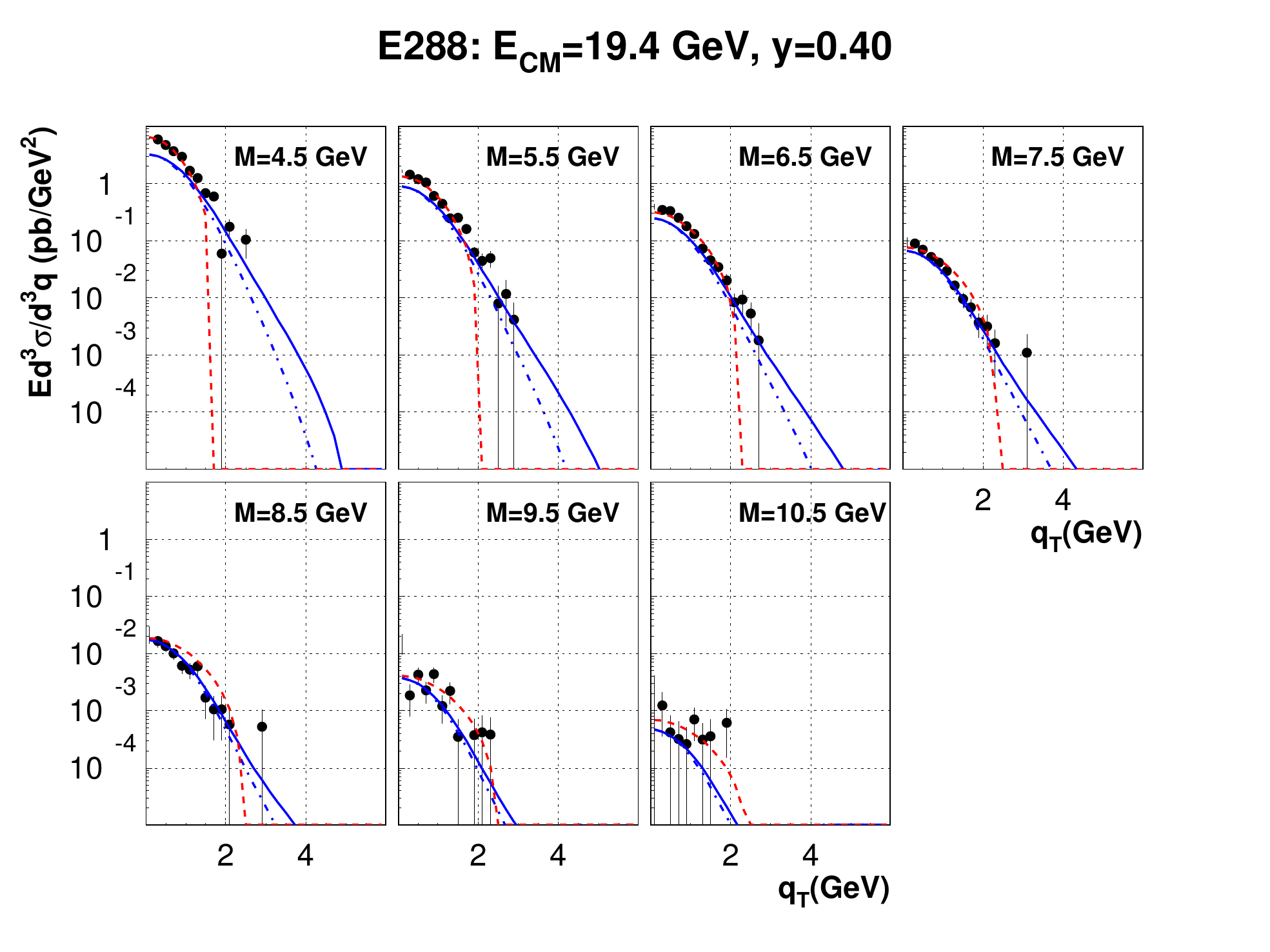}
\vskip -2mm
\includegraphics[width=0.7\textwidth]{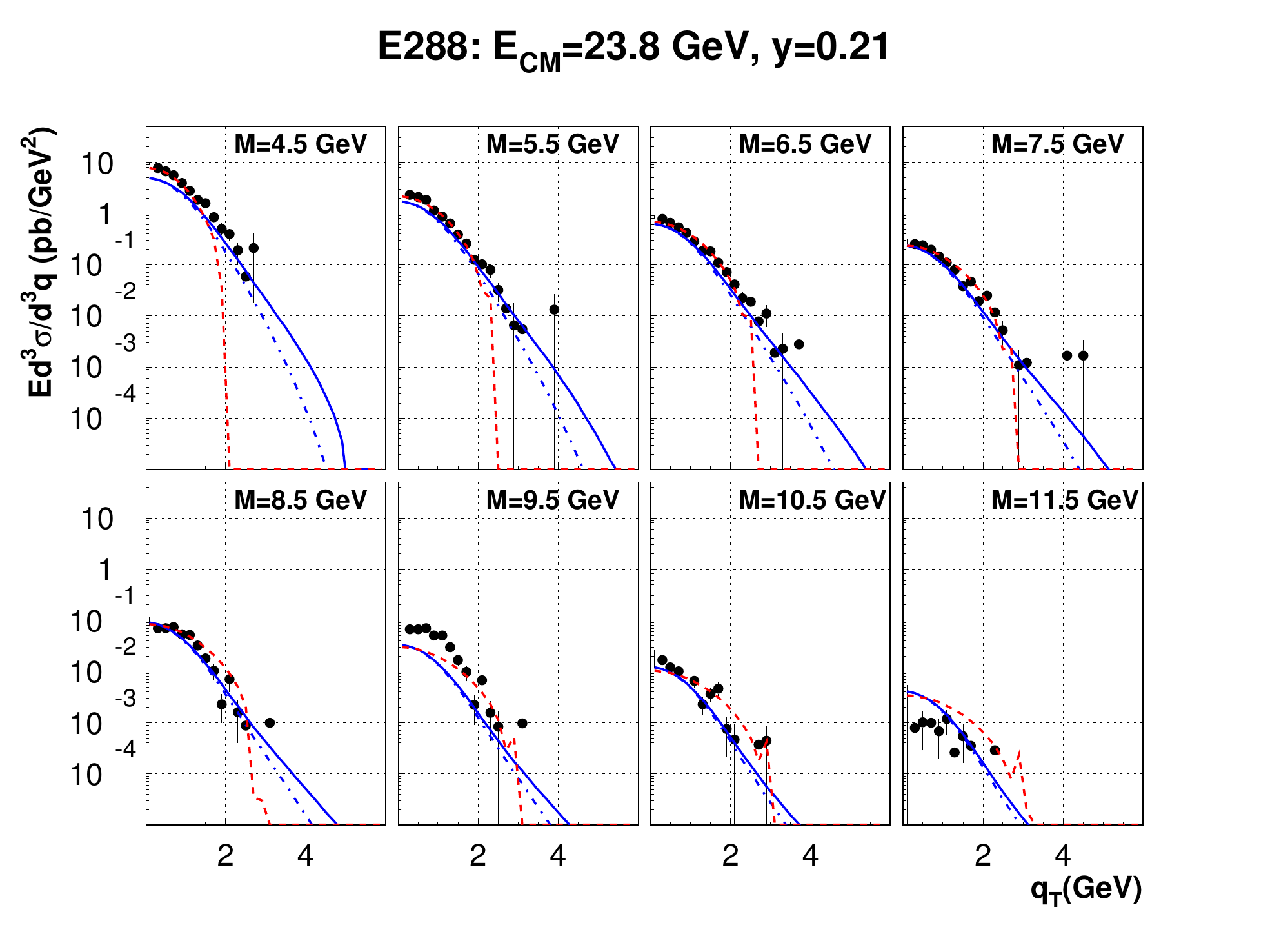}
\vskip -2mm
\includegraphics[width=0.7\textwidth]{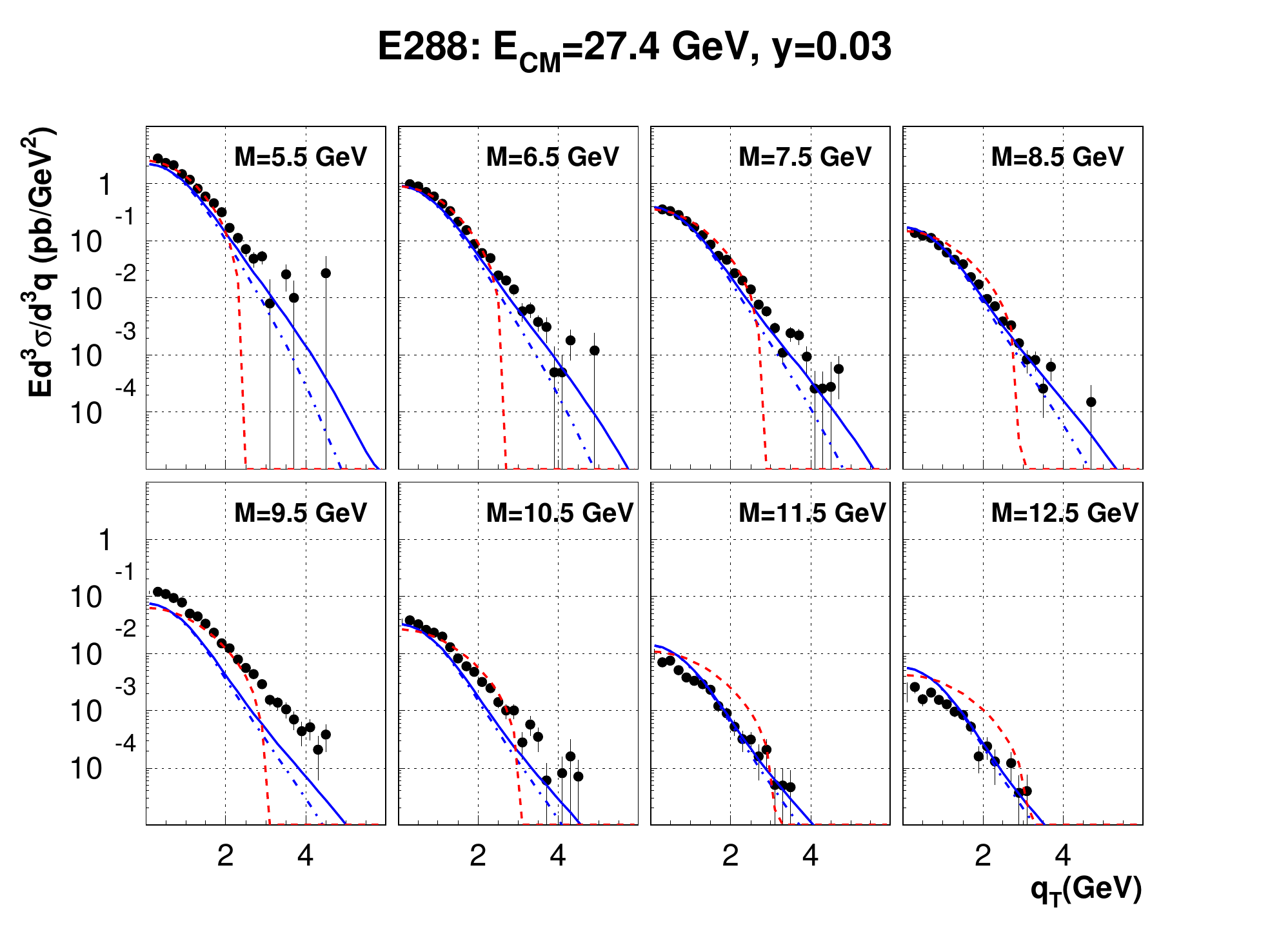}
\end{center}
\vskip -7mm
\caption{Transverse momentum dependence of DY cross section: data from fixed target experiment E288 \cite{Ito:1980ev} are compared with on-shell CCFM-K (blue solid), off-shell CCFM-K (blue dash-dotted) and CSS (red dashed) approaches.
}
\label{fig:e288}
\end{figure}

\begin{figure}[t]
\begin{center}
\includegraphics[width=0.8\textwidth]{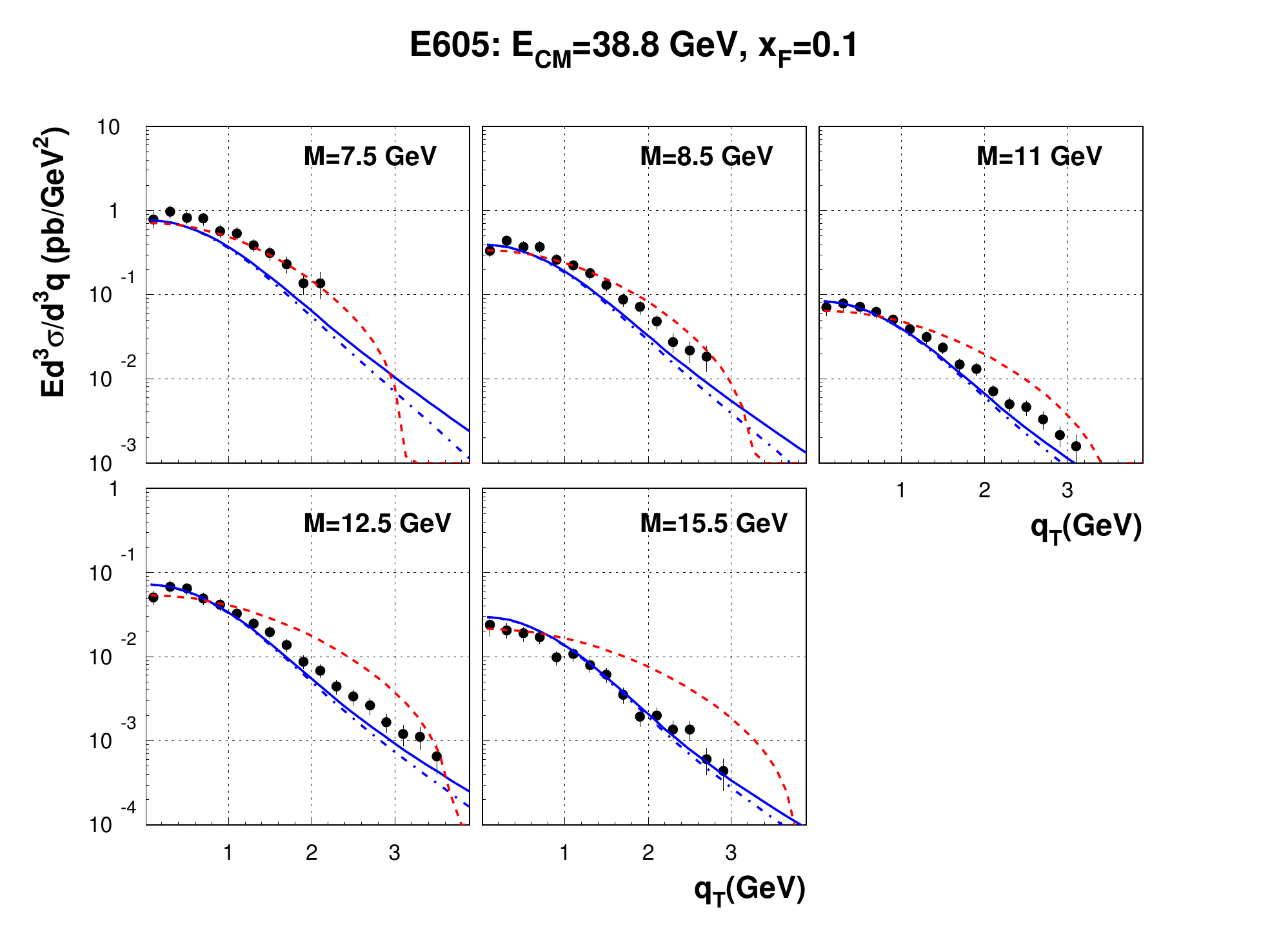}
\includegraphics[width=0.8\textwidth]{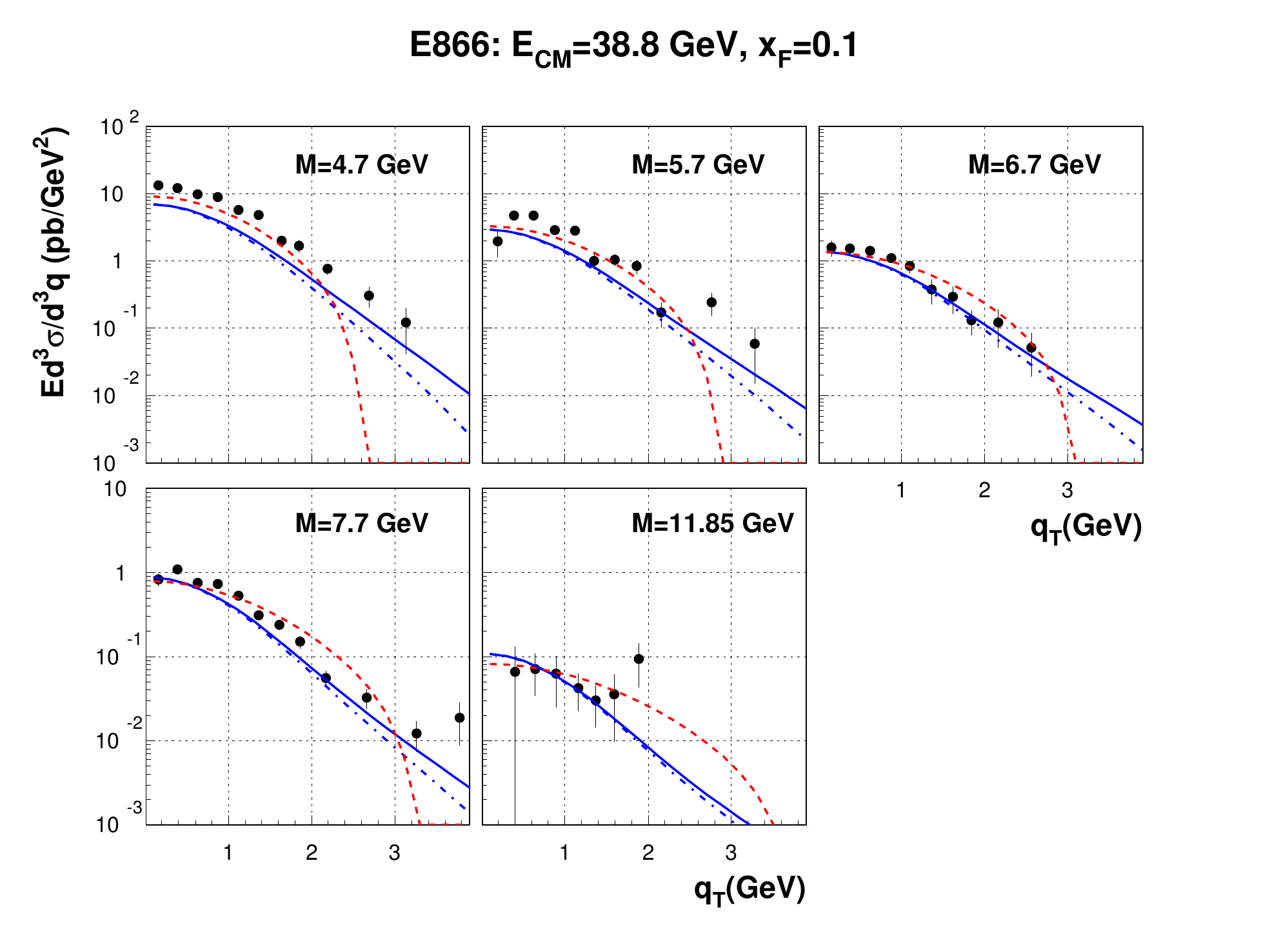}
\end{center}
\caption{Transverse momentum dependence of DY cross-section: data from fixed target experiments E605 \cite{Moreno:1990sf} (upper panels) and E866 \cite{Webb:2003bj} (lower panels) are compared with on-shell CCFM-K (blue solid), off-shell CCFM-K (blue dash-dotted) and CSS (red dashed) approaches.
}
\label{fig:e605_e866}
\end{figure}

\begin{figure}[t]
\vskip -15mm
\begin{center}
\includegraphics[width=0.72\textwidth]{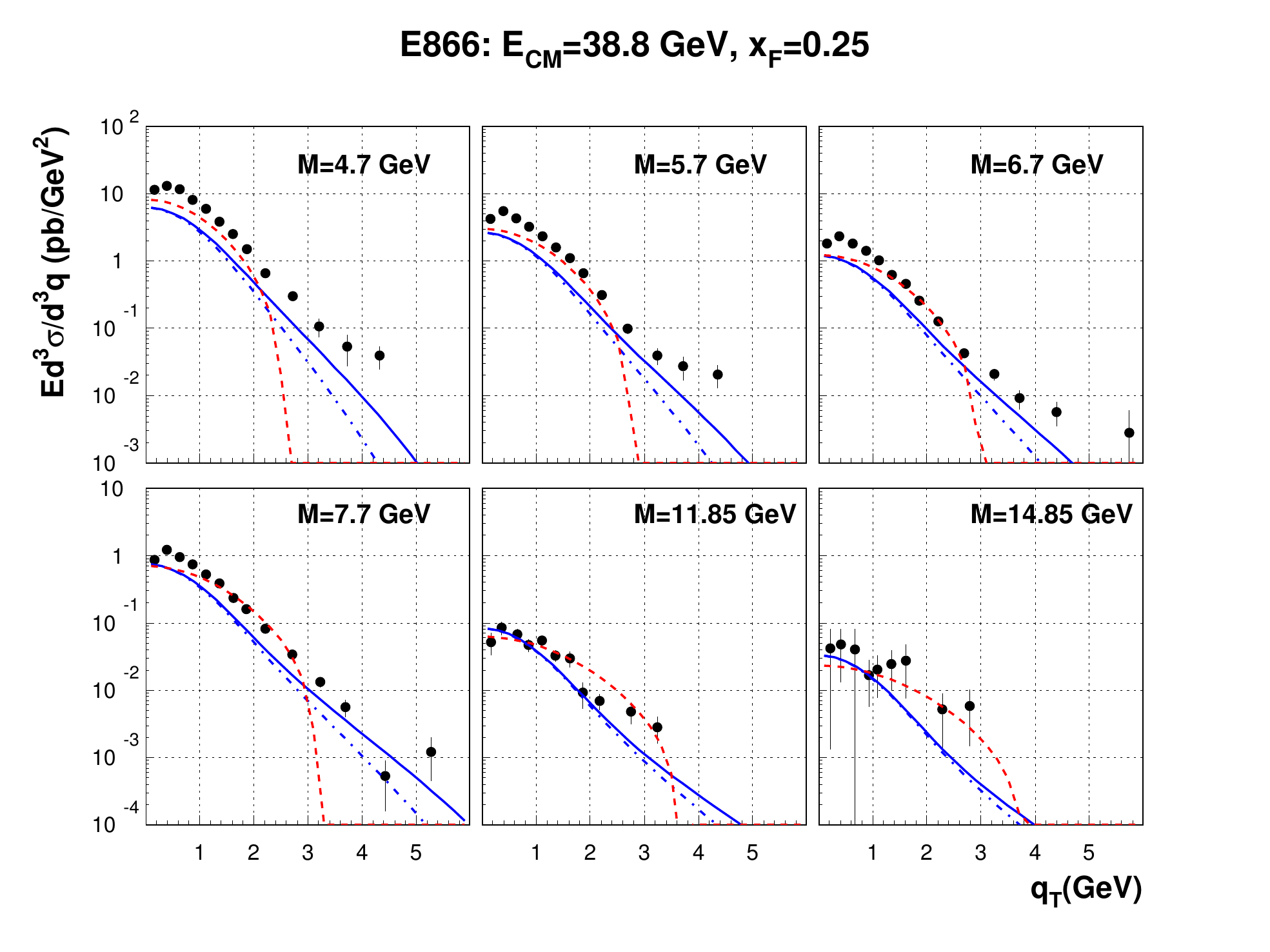}
\vskip -2mm
\includegraphics[width=0.72\textwidth]{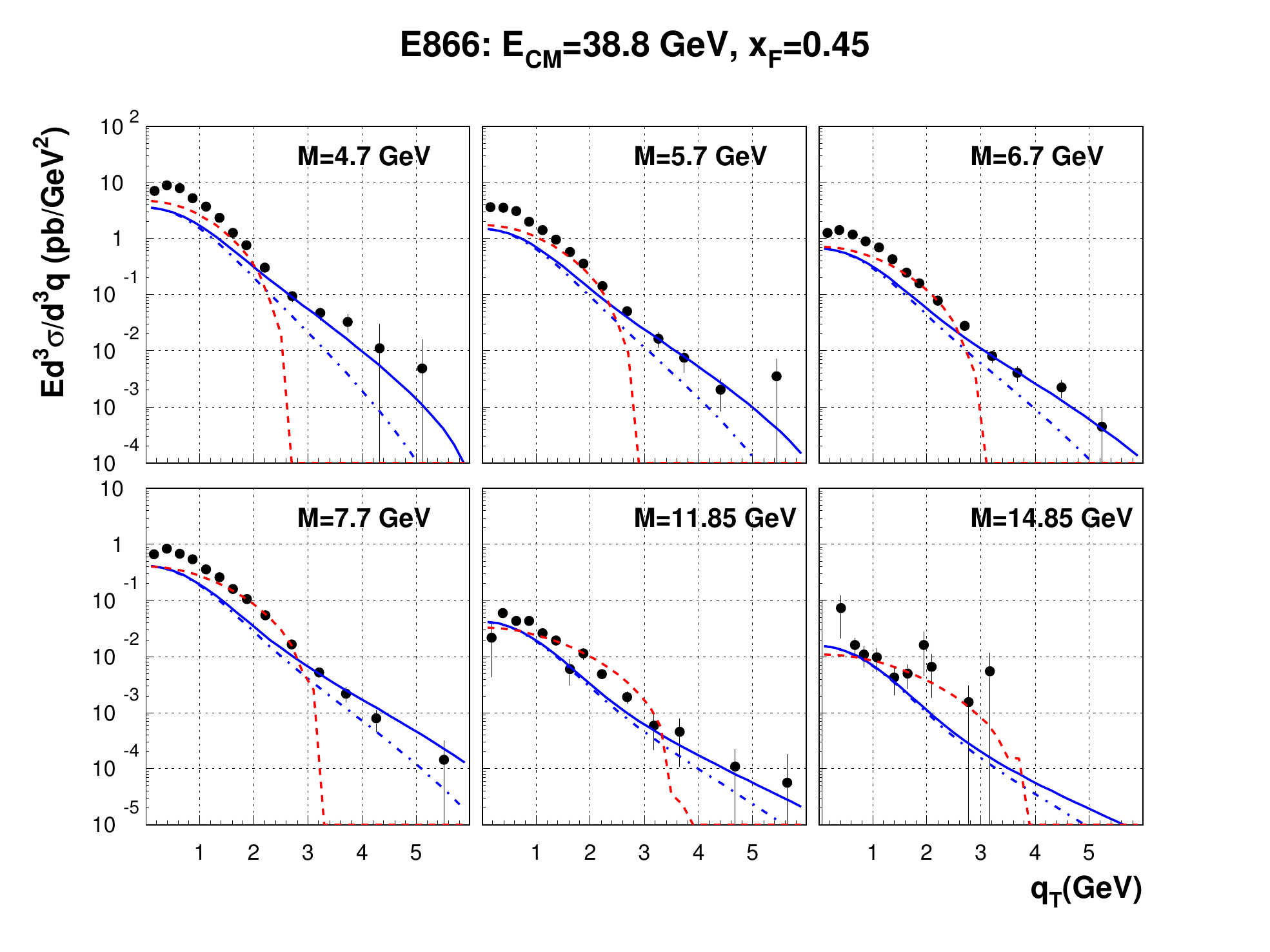}
\vskip -2mm
\includegraphics[width=0.72\textwidth]{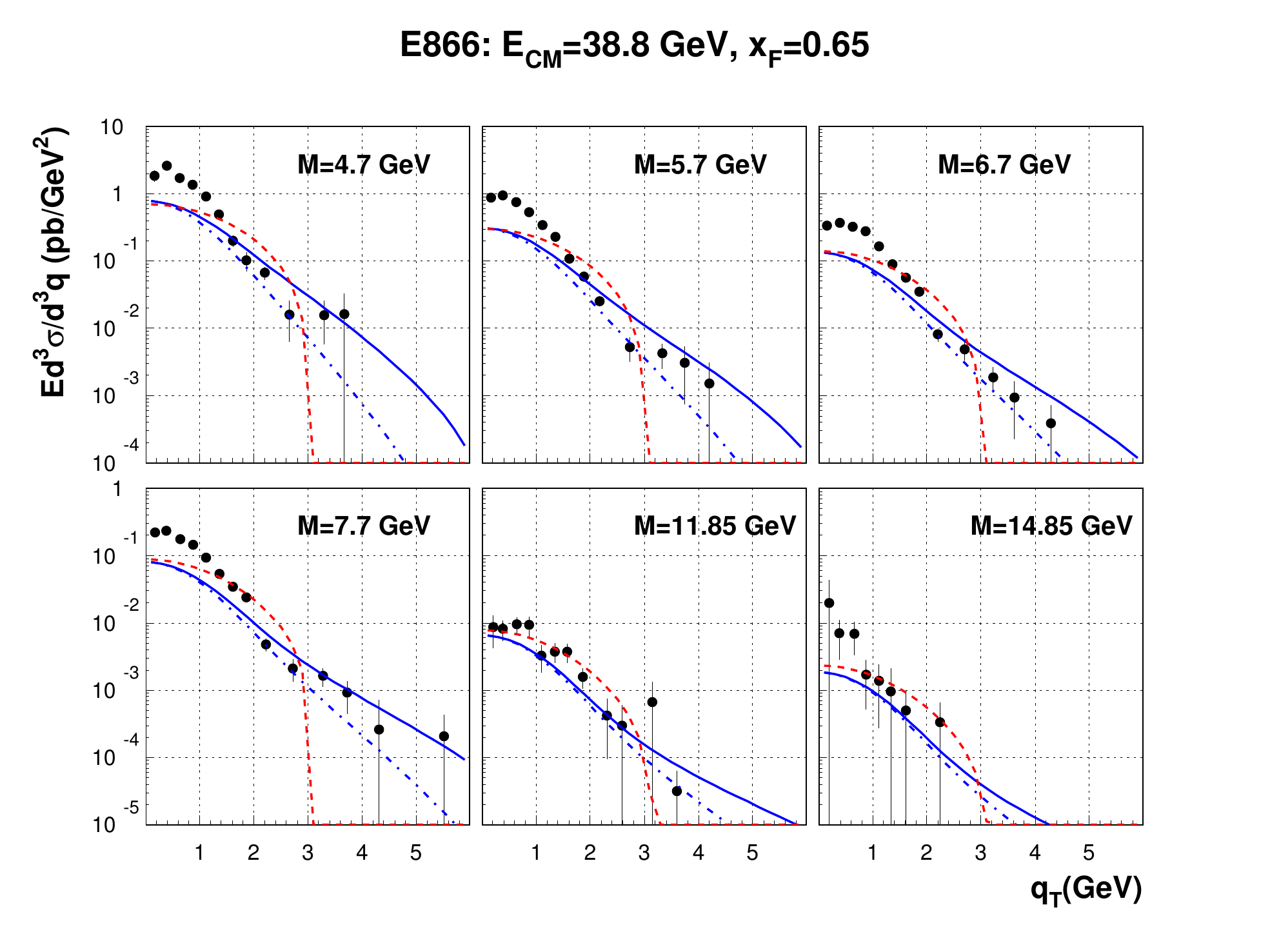}
\end{center}
\vskip -9mm
\caption{Transverse momentum dependence of DY cross-section: data from fixed target experiment E866 \cite{Webb:2003bj} are compared with on-shell CCFM-K (blue solid), off-shell CCFM-K (blue dash-dotted) and CSS (red dashed) approaches.
}
\label{fig:e866}
\end{figure}

\begin{figure}[t]
\begin{center}
\includegraphics[width=0.9\textwidth]{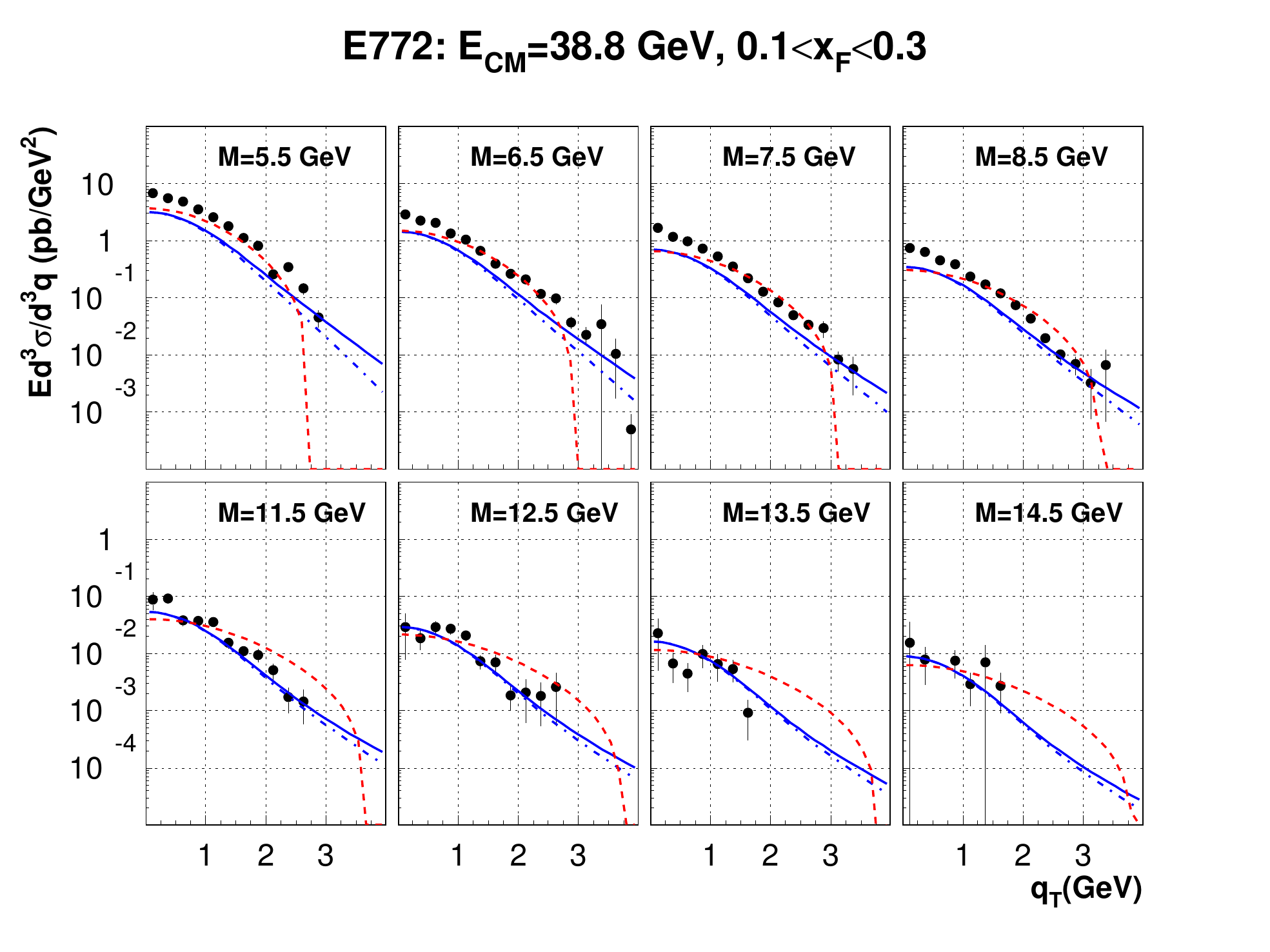}
\end{center}
\vskip -8mm
\caption{Transverse momentum dependence of DY cross-section: data from fixed target experiment E772 \cite{McGaughey:1994dx} are compared with on-shell CCFM-K (blue solid), off-shell CCFM-K (blue dash-dotted) and CSS (red dashed) approaches.
}
\label{fig:E772}
\end{figure}

\begin{figure}[t]
\begin{center}
\includegraphics[width=0.55\textwidth]{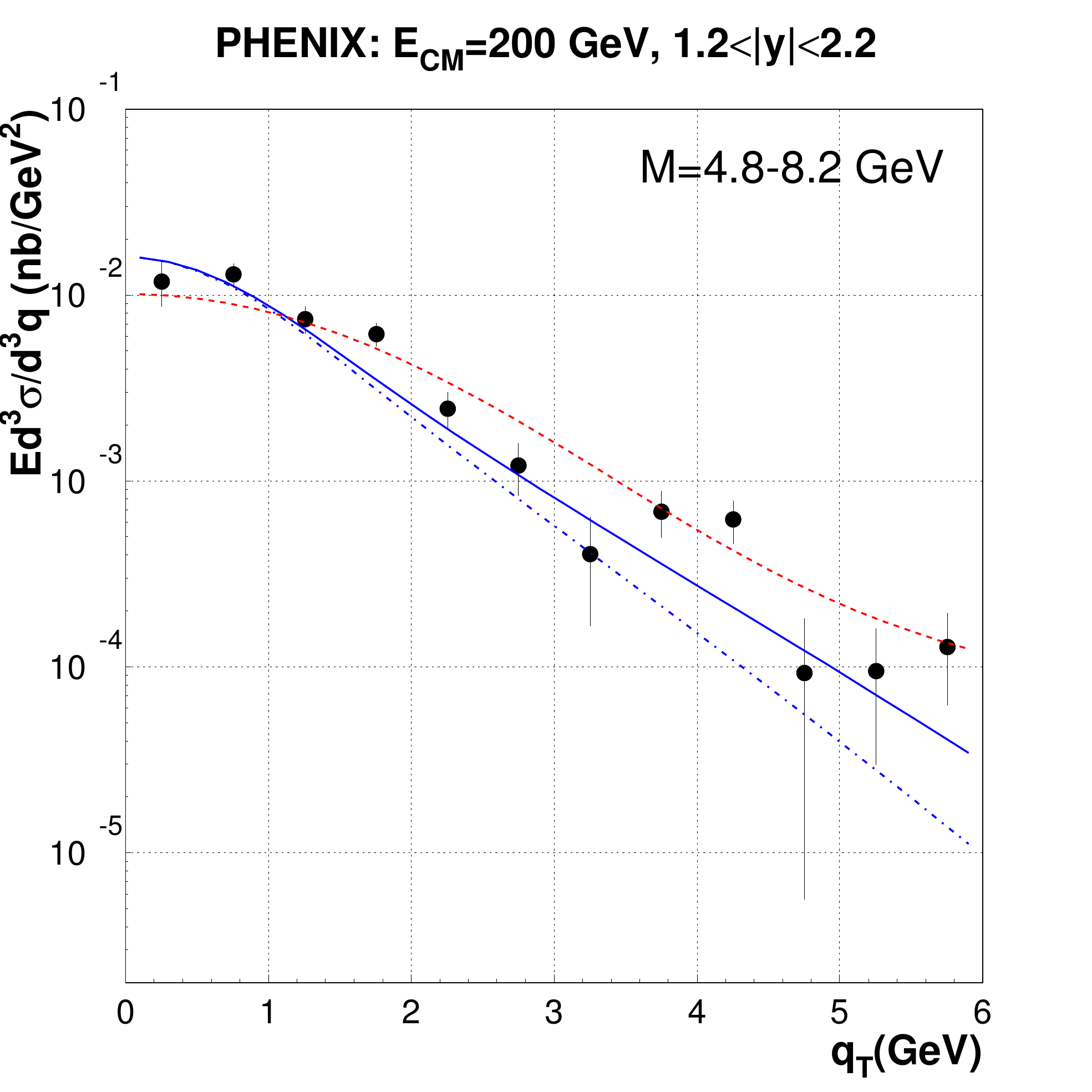}
\end{center}
\vskip -5mm
\caption{
Transverse momentum dependence of DY cross-section in proton-proton collisions: data from PHENIX \cite{Aidala:2018ajl} compared with on-shell CCFM-K (blue solid), off-shell CCFM-K (blue dash-dotted) and CSS (red dashed) approaches.
}
\label{fig:PHENIX}
\end{figure}

\bibliographystyle{JHEP}

\bibliography{mybib}

\end{document}